\documentclass[useAMS,usegraphicx,usenatbib]{mn2e}          

\usepackage{times}
\usepackage{flushend}

\newcommand{\Msun}{\ensuremath{\,{\rm M}_\odot}}                  
\newcommand{\Rsun}{\ensuremath{\,{\rm R}_\odot}}                  
\newcommand{\Teff}{\ensuremath{T_{\rm eff}}}                      
\newcommand{\logg}{\ensuremath{\log g}}                           
\newcommand{\Mjup}{\ensuremath{\,{\rm M}_{\rm Jup}}}              
\newcommand{\Rjup}{\ensuremath{\,{\rm R}_{\rm Jup}}}              
\newcommand{\Teq}{\ensuremath{T_{\rm eq}^{\,\prime}}}             
\newcommand{\Mearth}{\ensuremath{\,{\rm M}_\oplus}}               
\newcommand{\Porb}{\ensuremath{P_{\rm orb}}}                      
\newcommand{\kms}{\,km\,s$^{-1}$}                                 
\newcommand{\ms}{\,m\,s$^{-1}$}                                   
\newcommand{\mss}{\,m\,s$^{-2}$}                                  
\newcommand{\as}{\ensuremath{^{\prime\prime}}}                    
\newcommand{\am}{\ensuremath{^\prime}}                            
\newcommand{\FeH}{\ensuremath{\left[\frac{\rm Fe}{\rm H}\right]}} 
\newcommand{\pjup}{\ensuremath{\,\rho_{\rm Jup}}}                 
\newcommand{\psun}{\ensuremath{\,\rho_\odot}}                     
\newcommand{\chir}{\ensuremath{\chi_\nu^{\,2}}}                   
\newcommand{\mc}[1]{\multicolumn{2}{c}{#1}}
\newcommand{\mcc}[1]{\multicolumn{3}{c}{#1}}
\newcommand{\er}[3]{\ensuremath{#1^{+#2}_{-#3}}}
\newcommand{\erc}[3]{\mc{\ensuremath{#1^{+#2}_{-#3}}}}

\newcommand{\ercc}[3]{\mcc{\ensuremath{#1^{+#2}_{-#3}}}}
\newcommand{\ermcc}[5]{\mcc{\ensuremath{{#1\,^{+#2}_{-#3}}\,^{+#4}_{-#5}}}}
\newcommand{\reff}[1]{{#1}}

\title[WASP-22, WASP-41, WASP-42 and WASP-55]
      {High-precision photometry by telescope defocussing. VIII. WASP-22, WASP-41, WASP-42 and WASP-55%
      \thanks{Based on data collected by MiNDSTEp with the Danish 1.54\,m telescope at the ESO La Silla Observatory.}}

\author[Southworth et al.]
       {John Southworth\,$^{1}$, J.\ Tregloan-Reed\,$^{2}$, M.\ I.\ Andersen\,$^{3}$, S.\ {Calchi Novati}\,$^{4,5,6}$, S.\ Ciceri\,$^{7}$,
        \newauthor
        J.\ P.\ Colque\,$^{8}$, G.\ D'Ago\,$^{6}$, M.\ Dominik\,$^{9}$\thanks{Royal Society University Research Fellow}, D.\ F.\ Evans\,$^{1}$, S.-H.\ Gu\,$^{10,11}$, A.\ Herrera-Cordova\,$^{8}$,
        \newauthor
        T.\ C.\ Hinse\,$^{12}$, U.\ G.\ J{\o}rgensen\,$^{13}$, D.\ Juncher\,$^{13}$, M.\ Kuffmeier\,$^{13}$, L.\ Mancini\,$^{7,14}$,
        \newauthor
        N.\ Peixinho\,$^{8}$, A.\ Popovas\,$^{13}$, M.\ Rabus\,$^{15,7}$, J.\ Skottfelt\,$^{16,13}$, R.\ Tronsgaard\,$^{17}$,
        \newauthor
        E.\ Unda-Sanzana\,$^{8}$, X.-B.\ Wang\,$^{10,11}$, O.\ Wertz\,$^{18}$, K.\ A.\ Alsubai\,$^{19}$, J.\ M.\ Andersen\,$^{13,20}$,
        \newauthor
        V.\ Bozza\,$^{5,21}$, D.\ M.\ Bramich\,$^{19}$, M.\ Burgdorf\,$^{22}$, Y.\ Damerdji\,$^{18}$, C.\ Diehl\,$^{23,24}$,
        \newauthor
        A.\ Elyiv\,$^{25,18,26}$, R.\ {Figuera Jaimes}\,$^{9,27}$, T.\ Haugb{\o}lle\,$^{13}$, M.\ Hundertmark\,$^{13}$, N.\ Kains\,$^{28}$,
        \newauthor
        E.\ Kerins\,$^{29}$, H.\ Korhonen\,$^{30,13,3}$, C.\ Liebig\,$^{9}$, M.\ Mathiasen\,$^{13}$, M.\ T.\ Penny\,$^{31}$, S.\ Rahvar\,$^{32}$,
        \newauthor
        G.\ Scarpetta\,$^{6,5,21}$, R.\ W.\ Schmidt\,$^{23}$, C.\ Snodgrass\,$^{33}$, D.\ Starkey\,$^{9}$, J.\ Surdej\,$^{18}$, C.\ Vilela\,$^{1}$,
        \newauthor
        C.\ von Essen\,$^{17}$, Y.\ Wang\,$^{10}$
        \\
        $^{1}$\,Astrophysics Group, Keele University, Staffordshire, ST5 5BG, UK \\
        $^{2}$\,NASA Ames Research Center, Moffett Field, CA 94035, USA \\
        $^{3}$\,Dark Cosmology Centre, Niels Bohr Institute, University of Copenhagen, Juliane Maries vej 30, 2100 Copenhagen {\O}, Denmark \\
        $^{4}$\,NASA Exoplanet Science Institute, MS 100-22, California Institute of Technology, Pasadena, CA 91125, US \\
        $^{5}$\,Dipartimento di Fisica ``E.R. Caianiello'', Universit\`a di Salerno, Via Giovanni Paolo II 132, 84084, Fisciano (SA), Italy \\
        $^{6}$\,Istituto Internazionale per gli Alti Studi Scientifici (IIASS), 84019 Vietri Sul Mare (SA), Italy \\
        $^{7}$\,Max Planck Institute for Astronomy, K\"onigstuhl 17, 69117 Heidelberg, Germany \\
        $^{8}$\,Unidad de Astronom\'\i a, Facultad de Ciencias B\'asicas, Universidad de Antofagasta, Avenida U.\ de Antofagasta 02800, Antofagasta, Chile \\
        $^{9}$\,SUPA, University of St Andrews, School of Physics \& Astronomy, North Haugh, St Andrews, KY16 9SS, UK \\
        $^{10}$\,Yunnan Observatories, Chinese Academy of Sciences, Kunming 650011, China \\
        $^{11}$\,Key Laboratory for the Structure and Evolution of Celestial Objects, Chinese Academy of Sciences, Kunming 650011, China \\
        $^{12}$\,Korea Astronomy and Space Science Institute, Daejeon 305-348, Republic of Korea \\
        $^{13}$\,Niels Bohr Institute \& Centre for Star and Planet Formation, University of Copenhagen, {\O}ster Voldgade 5, 1350 Copenhagen K, Denmark \\
        $^{14}$\,INAF -- Osservatorio Astrofisico di Torino, via Osservatorio 20, 10025, Pino Torinese, Italy \\
        $^{15}$\,Instituto de Astrof{\'\i}sica, Facultad de F{\'\i}sica, Pontificia Universidad Cat\'olica de Chile, Av.\ Vicu\~na Mackenna 4860, 7820436 Macul, Santiago, Chile \\
        $^{16}$\,Centre of Electronic Imaging, Department of Physical Sciences, The Open University, Milton Keynes, MK7 6AA, UK \\
        $^{17}$\,Stellar Astrophysics Centre (SAC), Department of Physics and Astronomy, Aarhus University, Ny Munkegade 120, DK-8000 Aarhus C, Denmark \\
        $^{18}$\,Institut d'Astrophysique et de G\'eophysique, Universit\'e de Li\`ege, 4000 Li\`ege, Belgium \\
        $^{19}$\,Qatar Environment and Energy Research Institute (QEERI), HBKU, Qatar Foundation, PO Box 5825, Doha, Qatar \\
        $^{20}$\,Department of Astronomy, Boston University, 725 Commonwealth Avenue, Boston, MA 02215, USA \\
        $^{21}$\,Istituto Nazionale di Fisica Nucleare, Sezione di Napoli, 80126 Napoli, Italy \\
        $^{22}$\,Universit\"at Hamburg, Meteorologisches Institut, Bundesstra{\ss}e 55, 20146 Hamburg, Germany \\
        $^{23}$\,Astronomisches Rechen-Institut, Zentrum f\"ur Astronomie, Universit\"at Heidelberg, M\"onchhofstra{\ss}e 12-14, 69120 Heidelberg, Germany \\
        $^{24}$\,Hamburger Sternwarte, Universit\"at Hamburg, Gojenbergsweg 112, 21029 Hamburg, Germany \\
        $^{25}$\,Dipartimento di Fisica e Astronomia, Universit\`a di Bologna, Viale Berti Pichat 6/2, I-40127  Bologna, Italy \\
        $^{26}$\,Main Astronomical Observatory, Academy of Sciences of Ukraine, vul.\ Akademika Zabolotnoho 27, 03680 Kyiv, Ukraine \\
        $^{27}$\,European Southern Observatory, Karl-Schwarzschild-Stra{\ss}e 2, 85748 Garching bei M\"unchen, Germany \\
        $^{28}$\,Space Telescope Science Institute, 3700 San Martin Drive, Baltimore, MD 21218, USA \\
        $^{29}$\,Jodrell Bank Centre for Astrophysics, University of Manchester, Oxford Road, Manchester M13 9PL, UK \\
        $^{30}$\,Finnish Centre for Astronomy with ESO (FINCA), University of Turku, V{\"a}is{\"a}l{\"a}ntie 20, FI-21500 Piikki{\"o}, Finland \\
        $^{31}$\,Department of Astronomy, Ohio State University, 140 W. 18th Ave., Columbus, OH 43210, USA \\
        $^{32}$\,Department of Physics, Sharif University of Technology, P.\,O.\,Box 11155-9161 Tehran, Iran \\
        $^{33}$\,Planetary and Space Sciences, Department of Physical Sciences, The Open University, Milton Keynes, MK7 6AA, UK
        }

\begin{document} \maketitle 

\clearpage

\begin{abstract}
We present 13 high-precision and four additional light curves of four bright southern-hemisphere transiting planetary systems: WASP-22, WASP-41, WASP-42 and WASP-55. In the cases of WASP-42 and WASP-55, these are the first follow-up observations since their discovery papers. We present refined measurements of the physical properties and orbital ephemerides of all four systems. No indications of transit timing variations were seen. All four planets have radii inflated above those expected from theoretical models of gas-giant planets; WASP-55\,b is the most discrepant with a mass of 0.63\Mjup\ and a radius of 1.34\Rjup. WASP-41 shows brightness anomalies during transit due to the planet occulting spots on the stellar surface. Two anomalies observed 3.1\,d apart are very likely due to the same spot. We measure its change in position and determine a rotation period for the host star of $18.6 \pm 1.5$\,d, in good agreement with a published measurement from spot-induced brightness modulation, and a sky-projected orbital obliquity of $\lambda = 6 \pm 11^\circ$. We conclude with a compilation of obliquity measurements from spot-tracking analyses and a discussion of this technique in the study of the orbital configurations of hot Jupiters.
\end{abstract}



\section{Introduction}                                                                                                              \label{sec:intro}

Of the over 1200 transiting extrasolar planets (TEPs) now known\footnote{See Transiting Extrasolar Planet Catalogue (TEPCat; \citealt{Me11mn}): {\tt http://www.astro.keele.ac.uk/jkt/tepcat/}}, the short-period gas-giant planets are of particular interest. These `hot Jupiters' are the easiest to find due to their deep transits and high orbital frequency, are the most amenable to detailed characterisation due to their large masses and radii, and have highly irradiated and often rarefied atmospheres in which many physical phenomena are observable.

Most of the transiting hot Jupiters have been discovered by ground-based surveys studying bright stars. The brightness of the host stars is also extremely helpful in further characterisation of these objects via transmission spectroscopy and orbital obliquity studies. We are therefore undertaking a project to study TEPs orbiting bright host stars visible from the Southern hemisphere. Here we present transit light curves of four targets discovered by the WASP project \citep{Pollacco+06pasp} and measure their physical properties and orbital ephemerides to high precision.

WASP-22 was discovered by \citet{Maxted+10aj}, who found it to be a low-density planet (mass 0.56\Mjup, radius 1.12\Rjup) orbiting a $V=11.7$ solar-type star every 3.53\,d. A linear trend in the radial velocities (RVs) was noticed and attributed to the presence of a third body in the system, which could be an M-dwarf, white dwarf or second planet. The trend in the RVs has been confirmed by \citet{Knutson+14apj}, who measured the change in the systemic velocity of the system to be $\dot\gamma = \er{21.3}{2.8}{2.7}$\,m\,s$^{-1}$\,yr$^{-1}$. \citet{Anderson+11aa} measured the projected orbital obliquity of the system to be $\lambda = 22^\circ \pm 16^\circ$ via the Rossiter-McLaughlin effect.

WASP-41 was announced by \citet{Maxted+11pasp} to be a hot Jupiter of mass 0.94\Mjup, radius 1.06\Rjup, and orbital period $\Porb = 3.05$\,d. Its host is a $V = 11.6$ G8\,V star showing magnetic activity indicative of a young age, and rotational modulation on a period of $18.41 \pm 0.05$\,d. \citet{Neveu+15xxx} obtained further spectroscopic RV measurements from which they measured $\lambda = \er{29}{10}{14}$\degr\ and detected a third object in the system with $\Porb = 421 \pm 2$\,d and a minimum mass of $3.18 \pm 0.20$\Mjup.

WASP-42 was discovered by \citet{Lendl+12aa} and is a low-density planet (mass 0.50\Mjup, radius 1.12\Rjup) orbiting a $V=12.6$ star of spectral type K1\,V every 4.98\,d. An orbital eccentricity of $e = 0.060 \pm 0.0013$ was found by these authors, which is small but significant \citep{LucySweeney71aj}. No other study of the WASP-42 system has been published.

WASP-55 was one of a batch of new TEPs announced by \citet{Hellier+12mn} and is the lowest-density of the four planets considered here, with a mass of 0.50\Mjup\ and radius of 1.30\Rjup. Its host is a G1\,V star with a slightly sub-solar metallicity, and the \Porb\ of the system is 4.47\,d. No other study of the WASP-55 system has been published, but it was a target in Field 6 of the K2 mission \citep{Howell+14pasp} and these observations will soon be available.


\section{Observations and data reduction}                                                                                             \label{sec:obs}

\begin{figure*} \includegraphics[width=\textwidth,angle=0]{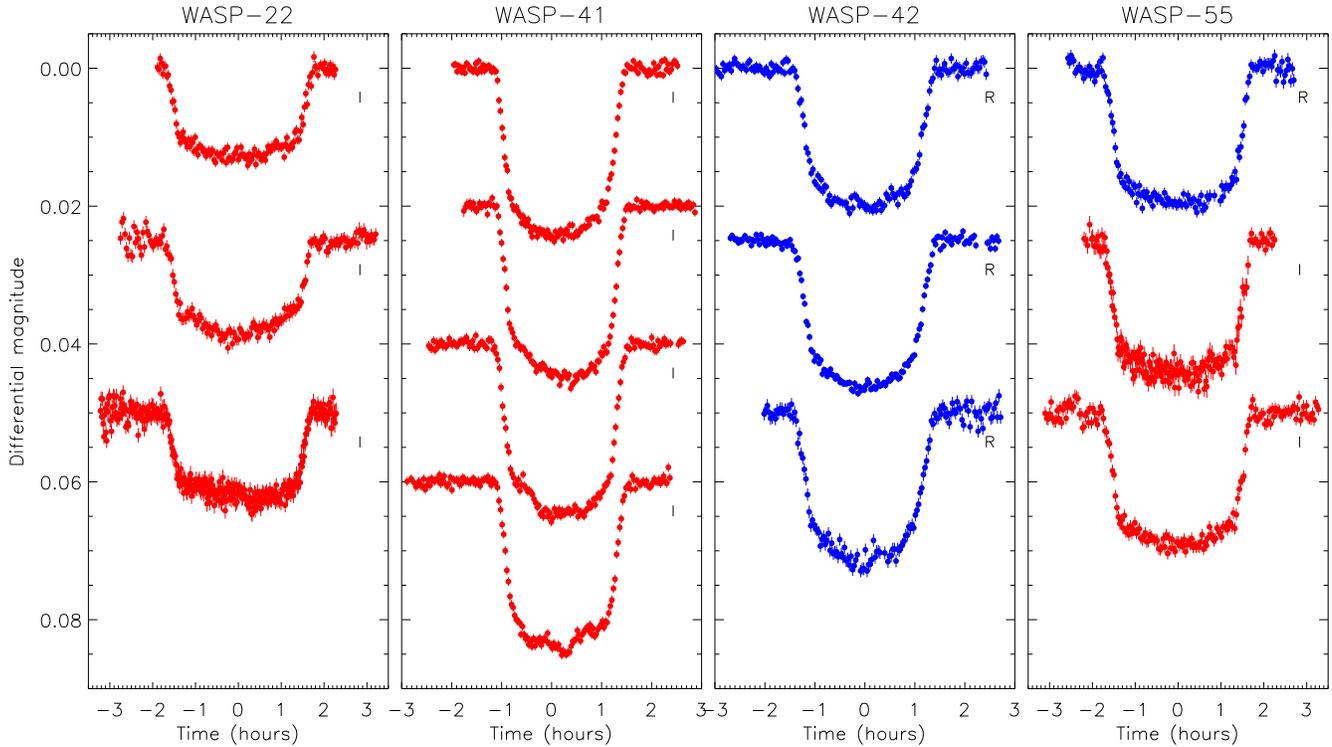}
\caption{\label{fig:lcall} DFOSC light curves presented in this work, in the
order they are given in Table\,\ref{tab:obslog}. Times are given relative
to the midpoint of each transit, and the filter used is indicated. Blue
and red filled circles represent observations through the Bessell $R$
and $I$ filters, respectively.} \end{figure*}

\begin{table*} \centering
\caption{\label{tab:obslog} Log of the observations presented in this work. $N_{\rm obs}$ is the number
of observations, $T_{\rm exp}$ is the exposure time, $T_{\rm dead}$ is the dead time between exposures,
`Moon illum.' is the fractional illumination of the Moon at the midpoint of the transit, given in italics
if the Moon was down at that time, and $N_{\rm poly}$ is the order of the polynomial fitted to the
out-of-transit data. The aperture radii refer to the target aperture, inner sky and outer sky, respectively.}
\setlength{\tabcolsep}{4pt}
\begin{tabular}{llcccccccccccc} \hline
Target  & Tele- & Date of   & Start time & End time &$N_{\rm obs}$ & $T_{\rm exp}$ & $T_{\rm dead}$ & Filter & Airmass &  Moon  & Aperture   & $N_{\rm poly}$ & Scatter \\
        & scope & first obs &    (UT)    &   (UT)   &              & (s)           & (s)            &        &         & illum. & radii (px) &                & (mmag)  \\
\hline
WASP-22 & 84\,cm & 2011 12 01 & 00:21 & 05:56 & 353 &    45   & 12 & none & 1.38 $\to$ 1.00 $\to$ 1.19 &      0.365  & 17 65 100 & 2 & 1.680 \\
WASP-22 & 84\,cm & 2012 01 23 & 00:54 & 04:53 & 228 &    45   & 10 & none & 1.02 $\to$ 1.02 $\to$ 2.46 & {\it 0.002} & 15 38  60 & 2 & 1.646 \\
WASP-22 & Danish & 2012 09 19 & 05:34 & 09:43 & 133 &   100   & 13 &  $I$ & 1.25 $\to$ 1.00 $\to$ 1.06 & {\it 0.137} & 18 32  50 & 1 & 0.703 \\ 
WASP-22 & Danish & 2013 09 25 & 03:10 & 09:08 & 172 &   100   & 25 &  $I$ & 2.21 $\to$ 1.00 $\to$ 1.05 &      0.680  & 18 25  55 & 1 & 0.913 \\ 
WASP-22 & Danish & 2015 09 05 & 04:38 & 10:08 & 369 & 35--45  & 12 &  $I$ & 2.11 $\to$ 1.00 $\to$ 1.03 &      0.510  & 14 20  40 & 1 & 1.087 \\[3pt] 
WASP-41 & 84\,cm & 2011 02 07 & 06:39 & 09:44 & 182 &    50   & 16 &  $R$ & 1.07 $\to$ 1.00 $\to$ 1.07 & {\it 0.161} & 25 50 100 & 2 & 2.942 \\
WASP-41 & 84\,cm & 2012 01 21 & 06:15 & 09:40 & 102 &    90   &  7 & none & 1.33 $\to$ 1.01 $\to$ 1.01 & {\it 0.045} & 27 54 108 & 2 & 1.729 \\
WASP-41 & Danish & 2014 05 31 & 00:35 & 05:03 & 155 & 80--100 & 13 &  $I$ & 1.01 $\to$ 1.00 $\to$ 1.71 & {\it 0.055} & 22 30  55 & 1 & 0.571 \\ 
WASP-41 & Danish & 2015 05 10 & 22:51 & 03:28 & 148 &   100   & 13 &  $I$ & 1.37 $\to$ 1.00 $\to$ 1.05 & {\it 0.548} & 22 28  50 & 2 & 0.596 \\ 
WASP-41 & Danish & 2015 05 13 & 23:25 & 04:30 & 159 &   100   & 13 &  $I$ & 1.19 $\to$ 1.00 $\to$ 1.19 & {\it 0.214} & 17 27  45 & 2 & 0.646 \\ 
WASP-41 & Danish & 2015 05 17 & 00:14 & 05:30 & 166 &   100   & 13 &  $I$ & 1.06 $\to$ 1.00 $\to$ 1.50 & {\it 0.015} & 22 28  50 & 1 & 0.646 \\[3pt] 
WASP-42 & Danish & 2013 05 25 & 00:51 & 06:28 & 164 &   100   & 20 &  $R$ & 1.04 $\to$ 1.03 $\to$ 2.10 &      1.000  & 20 28  45 & 2 & 0.673 \\ 
WASP-42 & Danish & 2013 06 18 & 23:11 & 04:33 & 168 &   100   & 16 &  $R$ & 1.04 $\to$ 1.03 $\to$ 1.92 &      0.745  & 19 27  50 & 1 & 0.501 \\ 
WASP-42 & Danish & 2013 06 28 & 23:00 & 03:43 & 143 &   100   & 15 &  $R$ & 1.04 $\to$ 1.03 $\to$ 1.82 & {\it 0.623} & 22 30  55 & 1 & 0.924 \\[3pt] 
WASP-55 & Danish & 2013 05 04 & 02:26 & 07:42 & 152 & 90--98  & 25 &  $R$ & 1.05 $\to$ 1.02 $\to$ 1.94 & {\it 0.314} & 17 42  80 & 1 & 0.815 \\ 
WASP-55 & Danish & 2014 06 18 & 22:56 & 03:26 & 206 & 50--110 & 11 &  $I$ & 1.10 $\to$ 1.02 $\to$ 1.35 & {\it 0.583} & 13 42  80 & 1 & 1.144 \\ 
WASP-55 & Danish & 2015 04 23 & 01:06 & 07:29 & 184 &   100   & 25 &  $I$ & 1.41 $\to$ 1.02 $\to$ 1.45 & {\it 0.231} & 16 26  50 & 1 & 0.899 \\ 
\hline \end{tabular} \end{table*}

\begin{figure} \includegraphics[width=\columnwidth,angle=0]{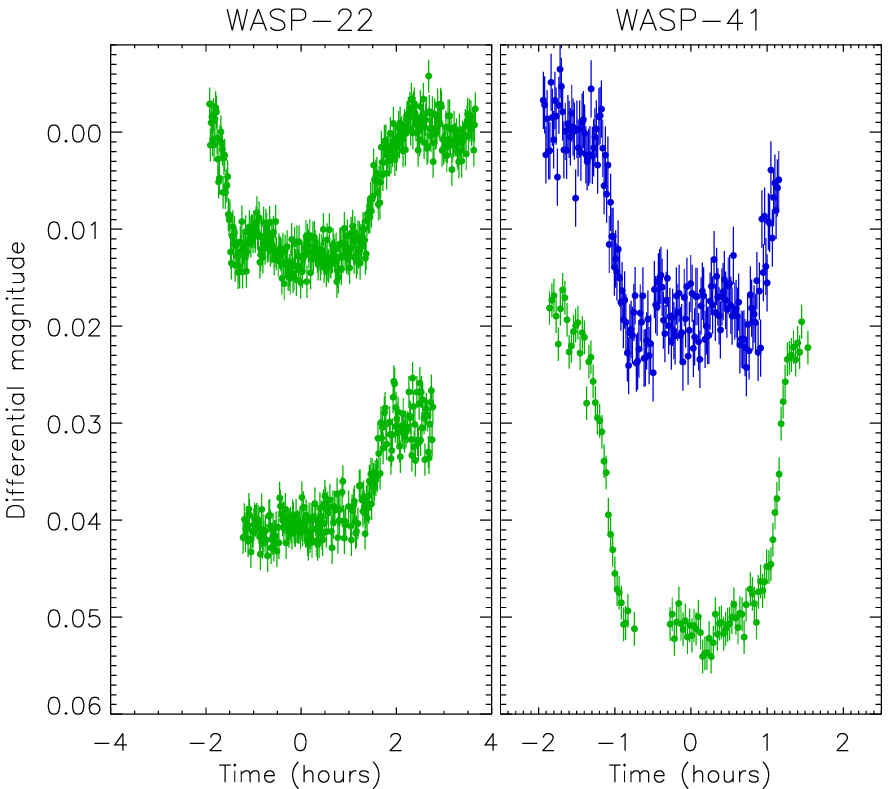}
\caption{\label{fig:lc84} 84\,cm telescope light curves presented in this
work, in the order they are given in Table\,\ref{tab:obslog}. Times are
given relative to the midpoint of each transit, and the filter used is
indicated. Blue and green filled circles represent observations through
the $R$ filter and without filter, respectively.} \end{figure}

We observed a total of 13 transits with the DFOSC (Danish Faint Object Spectrograph and Camera) instrument installed on the 1.54\,m Danish Telescope at ESO La Silla, Chile. DFOSC has a field of view of 13.7\am$\times$13.7\am\ at a plate scale of 0.39\as\,pixel$^{-1}$. We defocussed the telescope in order to improve the precision and efficiency of our observations \citep{Me+09mn}. The CCD was windowed during some observing sequences in order to shorten the readout time, and no binning was used. In most cases the night was photometric; observations taken through thin cloud were carefully checked and rejected if their reliability was questionable. The data were taken through either a Bessell $R$ or Bessell $I$ filter. An observing log is given in Table\,\ref{tab:obslog} and the final light curves are plotted in Fig.\,\ref{fig:lcall}. All observations were taken after the upgrade of the telescope and CCD controller in 2011 \citep{Me+14mn}.

We reduced the data using the {\sc defot} pipeline \citep[see][and references therein]{Me+14mn}, which in turn uses the {\sc idl}\footnote{{\tt http://www.exelisvis.co.uk/ProductsServices/IDL.aspx}} implementation of the {\sc aper} routine from {\sc daophot} \citep{Stetson87pasp} contained in the NASA {\sc astrolib} library\footnote{{\tt http://idlastro.gsfc.nasa.gov/}}. For each dataset, the apertures were placed by hand \reff{on a reference image. Shifts between the individual images and the reference image were measured by cross-correlation, and applied to the aperture positions.} The radii of the object aperture and sky annulus were chosen to minimise the scatter in the final light curve (see Table\,\ref{tab:obslog}). The science images were not calibrated using bias or flat-field frames as these tend to have little effect on the final light curves beyond a slight increase in the scatter of the datapoints.

We observed two transits of WASP-22 and two transits of WASP-41 using the 84\,cm telescope at Observatorio Cerro Armazones in Antofagasta, Chile (currently decommissioned). Three transits were observed using an SBIG ST-10 CCD camera, giving a field of view of 15.6\am$\times$10.5\am\ at a plate scale of 0.43\as\,pixel$^{-1}$, and the first transit of WASP-41 was monitored using an SBIG STL CCD camera with a field of view of 29.3\am$\times$19.5\am\ at a plate scale of 0.57\as\,pixel$^{-1}$. We defocussed the telescope and windowed the CCDs, and observed unfiltered except for the first transit of WASP-41 which was seen through an $R$ filter. Data reduction was performed using a custom pipeline based on Starlink routines, including calibration with dark frames but not flat-fields. The {\sc photom} package \citep{Eaton++99} was used to perform aperture photometry, and the apertures were placed by hand. A growth-curve analysis was performed for each dataset in order to find the aperture size which gave the lowest scatter. The data are plotted in Fig.\,\ref{fig:lc84}

For all datasets, differential-magnitude light curves were generated for each target star versus an ensemble comparison star containing the weighted flux sum of \reff{the best three to five} comparison stars. A polynomial was also fitted to the observations outside transit and subtracted to rectify the final light curve to zero differential magnitude. In most cases a first-order polynomial was an adequate match to the slow brightness variations seen throughout the observing sequences, but in some cases a quadratic was required (see Table\,\ref{tab:obslog}). The coefficients of the polynomial and the weights of the comparison stars were simultaneously optimised to minimise the scatter in the datapoints outside eclipse.

Manual time checks were obtained for several frames and the FITS file timestamps were confirmed to be on the UTC system to within a few seconds. They were then converted to the BJD(TDB) timescale \citep{Eastman++10pasp}. The light curves are shown in Fig.\,\ref{fig:lcall}, and the reduced data (Table\,\ref{tab:lcdata}) will be made available at the CDS\footnote{{\tt http://vizier.u-strasbg.fr/}}.

\begin{table} \centering \caption{\label{tab:lcdata} The
first line of each of the light curves presented in this
work. The full dataset will be made available at the CDS.}
\setlength{\tabcolsep}{4pt}
\begin{tabular}{llccrr} \hline
Target & Tele- & Filter & BJD(TDB)   & Diff.\ mag. & Uncertainty \\
       & scope &        & $-$2400000 &             &             \\
\hline
WASP-22 & 84\,cm & none & 55896.520223 & $-$0.00291 & 0.00168 \\
WASP-22 & 84\,cm & none & 55949.539861 & $-$0.01180 & 0.00165 \\
WASP-22 & Danish & $I$  & 56189.736334 & $-$0.00015 & 0.00074 \\
WASP-22 & Danish & $I$  & 56560.637168 & $-$0.00034 & 0.00096 \\
WASP-22 & Danish & $I$  & 57270.696685 & $-$0.00032 & 0.00105 \\
WASP-41 & 84\,cm & $R$  & 55599.780710 & $-$0.00329 & 0.00294 \\
WASP-41 & 84\,cm & none & 55947.763186 & $-$0.00187 & 0.00173 \\
WASP-41 & Danish & $I$  & 56808.529719 & $-$0.00050 & 0.00054 \\
WASP-41 & Danish & $I$  & 57153.458491 &    0.00071 & 0.00089 \\
WASP-41 & Danish & $I$  & 57156.481701 &    0.00025 & 0.00068 \\
WASP-41 & Danish & $I$  & 57159.515981 & $-$0.00007 & 0.00065 \\
WASP-42 & Danish & $R$  & 56437.541226 &    0.00099 & 0.00061 \\
WASP-42 & Danish & $R$  & 56462.470118 & $-$0.00036 & 0.00049 \\
WASP-42 & Danish & $R$  & 56472.461607 &    0.00013 & 0.00097 \\
WASP-55 & Danish & $R$  & 56416.608745 & $-$0.00146 & 0.00083 \\
WASP-55 & Danish & $I$  & 56827.462644 & $-$0.00027 & 0.00105 \\
WASP-55 & Danish & $I$  & 57135.553466 &    0.00004 & 0.00094 \\
\hline \end{tabular} \end{table}

Finally, each of the light curves was fitted with the {\sc jktebop} code (see below) in order to determine the quality of fit and the times of midpoint of the transits. The errorbars for each dataset were rescaled to give a reduced $\chi^2$ of $\chir = 1.0$ versus the fitted model, necessary as the uncertainties from the {\sc aper} algorithm are often underestimated.


\section{Light curve analysis}                                                                                                         \label{sec:lc}

We modelled the light curves of the four targets using the {\it Homogeneous Studies} methodology (see \citealt{Me12mn} and references therein), which utilises the {\sc jktebop}\footnote{{\sc jktebop} is written in {\sc fortran77} and the source code is available at {\tt http://www.astro.keele.ac.uk/jkt/codes/jktebop.html}} code \citep[][and references therein]{Me13aa}. {\sc jktebop} represents the star and planet as spheres for the calculation of eclipse shapes and as biaxial spheroids for proximity effects.

The fitted parameters in our analysis were the fractional radii of the star and planet ($r_{\rm A}$ and $r_{\rm b}$), the orbital inclination ($i$), limb darkening coefficients, and a reference time of mid-transit. The fractional radii are the ratio between the true radii and the semimajor axis: $r_{\rm A,b}= \frac{R_{\rm A,b}}{a}$, and were expressed as their sum and ratio, $r_{\rm A} + r_{\rm b}$ and $k = \frac{r_{\rm b}}{r_{\rm A}}$, as these quantities are less strongly correlated. The orbital periods were fixed at the values found in Section\,\ref{sec:porb}. A polynomial of brightness versus time was applied to each transit light curve, with a polynomial order as given in Table\,\ref{tab:obslog}. \reff{This is not needed to fit the data, as the polynomial has already been removed at the data reduction stage, but is necessary to include the uncertainties of polynomial fit in the errorbars of the photometric parameters} \citep[see][]{Me+14mn}.

Limb darkening (LD) was incorporated into the photometric model using each of five LD laws \citep[see][]{Me08mn}, with the linear coefficients either fixed at theoretically predicted values\footnote{Theoretical LD coefficients were obtained by bilinear interpolation to the host star's \Teff\ and \logg\ using the {\sc jktld} code available from: {\tt http://www.astro.keele.ac.uk/jkt/codes/jktld.html}} or included as fitted parameters. We did not calculate fits for both LD coefficients in the four two-parameter laws as they are strongly correlated \citep{Me08mn,Carter+08apj}. The nonlinear coefficients were instead perturbed by $\pm$0.1 on a flat distribution during the error analysis simulations, in order to account for uncertainties in the theoretical coefficients.

All four targets have been observed in the HITEP high-resolution imaging campaign by \citet{Evans+15sub} using the Two Colour Imager (TCI) \citep{Skottfelt+15aa} to perform Lucky Imaging. No stars were found close enough to WASP-22, WASP-41 or WASP-42 to affect our photometry. However, one star was found at an angular distance of $4.345 \pm 0.010$\,arcsec from WASP-55, and was accounted for in the {\sc jktebop} model (see below).

Error estimates for the fitted parameters were obtained in several ways. We ran solutions using different LD laws, and also calculated errorbars using residual-permutation and Monte Carlo algorithms \citep{Me08mn}. The final value for each parameter is the unweighted mean of the four values from the solutions using the two-parameter LD laws. Its errorbar was taken to be the larger of the Monte-Carlo or residual-permutation alternatives, with an extra contribution to account for variations between solutions with the different LD laws. Tables of results for each light curve can be found in the Supplementary Information.

\begin{table*} \caption{\label{tab:lcfit} Parameters of the {\sc jktebop}
fits to the new and published light curves of the four planetary systems.}
\begin{tabular}{l l l r@{\,$\pm$\,}l r@{\,$\pm$\,}l r@{\,$\pm$\,}l r@{\,$\pm$\,}l r@{\,$\pm$\,}l}
\hline
System  & Source   & Filter  & \mc{$r_{\rm A}+r_{\rm b}$} & \mc{$k$} & \mc{$i$ ($^\circ$)} & \mc{$r_{\rm A}$} & \mc{$r_{\rm b}$} \\
\hline
WASP-22    & DFOSC       & $I$        & \erc{0.1284}{0.0039}{0.0017} & \erc{0.0996}{0.0013}{0.0012} & \erc{89.3}{1.0}{1.3} & \erc{0.1168}{0.0035}{0.0015} & \erc{0.01163}{0.00048}{0.00020} \\
WASP-22    & TRAPPIST    & $I$+$z$    & \erc{0.1327}{0.0083}{0.0059} & \erc{0.0951}{0.0015}{0.0013} & \erc{87.9}{2.1}{1.3} & \erc{0.1212}{0.0074}{0.0053} & \erc{0.01152}{0.00084}{0.00059} \\
WASP-22    & Euler       & $r$        & \erc{0.1404}{0.0121}{0.0090} & \erc{0.0982}{0.0041}{0.0049} & \erc{87.1}{2.6}{1.7} & \erc{0.1278}{0.0109}{0.0078} & \erc{0.01255}{0.00166}{0.00112} \\[3pt]
\multicolumn{3}{l}{\it Weighted mean} & \erc{0.1310}{0.0031}{0.0028} & \erc{0.0978}{0.0012}{0.0012} & \erc{88.6}{1.0}{1.0} & \erc{0.1193}{0.0027}{0.0026} & \erc{0.01172}{0.00039}{0.00025} \\
\hline
WASP-41    & DFOSC       & $I$        & 0.1143 & 0.0013 & 0.1362 & 0.0008 & 89.07 & 0.53 & 0.1006 & 0.0011 & 0.01369 & 0.00021 \\
WASP-41    & DFOSC       & $R$        & 0.1128 & 0.0015 & 0.1369 & 0.0013 & 89.62 & 1.06 & 0.0992 & 0.0013 & 0.01358 & 0.00027 \\
WASP-41    & TRAPPIST    & $I$+$z$    & 0.1176 & 0.0034 & 0.1378 & 0.0019 & 88.26 & 0.98 & 0.1034 & 0.0028 & 0.01424 & 0.00053 \\
WASP-41    & FTS         & $z$        & 0.1198 & 0.0054 & 0.1366 & 0.0018 & 87.63 & 0.80 & 0.1054 & 0.0047 & 0.01440 & 0.00078 \\[3pt]
\multicolumn{3}{l}{\it Weighted mean} & 0.1142 & 0.0009 & 0.1365 & 0.0006 & 88.70 & 0.39 & 0.1004 & 0.0008 & 0.01373 & 0.00015 \\
\hline
WASP-42    & DFOSC       & $R$        & 0.0851 & 0.0027 & 0.1296 & 0.0009 & 87.91 & 0.19 & 0.0753 & 0.0024 & 0.00976 & 0.00034 \\
WASP-42    & Euler       & $r$        & 0.0794 & 0.0047 & 0.1275 & 0.0047 & 88.72 & 1.01 & 0.0704 & 0.0040 & 0.00897 & 0.00071 \\
WASP-42    & TRAPPIST    & $I$+$z$    & 0.0829 & 0.0033 & 0.1284 & 0.0016 & 88.23 & 0.36 & 0.0735 & 0.0028 & 0.00943 & 0.00043 \\[3pt]
\multicolumn{3}{l}{\it Weighted mean} & 0.0834 & 0.0019 & 0.1293 & 0.0008 & 88.00 & 0.17 & 0.0739 & 0.0017 & 0.00955 & 0.00025 \\
\hline
WASP-55    & DFOSC       & $R$        & \erc{0.1025}{0.0030}{0.0009} & \erc{0.1253}{0.0014}{0.0010} & \erc{89.83}{0.57}{1.20} & \erc{0.0911}{0.0026}{0.0008} & \erc{0.01141}{0.00044}{0.00012} \\
WASP-55    & DFOSC       & $I$        & \erc{0.1028}{0.0024}{0.0008} & \erc{0.1236}{0.0009}{0.0008} & \erc{89.73}{0.59}{0.93} & \erc{0.0915}{0.0021}{0.0007} & \erc{0.01130}{0.00033}{0.00012} \\
WASP-55    & Euler       & $r$        & \erc{0.1102}{0.0064}{0.0064} & \erc{0.1274}{0.0024}{0.0025} & \erc{87.79}{0.98}{0.86} & \erc{0.0978}{0.0055}{0.0054} & \erc{0.01246}{0.00085}{0.00088} \\[3pt]
\multicolumn{3}{l}{\it Weighted mean} & \erc{0.1033}{0.0018}{0.0010} & \erc{0.1246}{0.0007}{0.0007} & \erc{89.05}{0.59}{0.59} & \erc{0.0918}{0.0015}{0.0009} & \erc{0.01143}{0.00025}{0.00013} \\
\hline \end{tabular} \end{table*}

\subsection{WASP-22}

\begin{figure} \includegraphics[width=\columnwidth,angle=0]{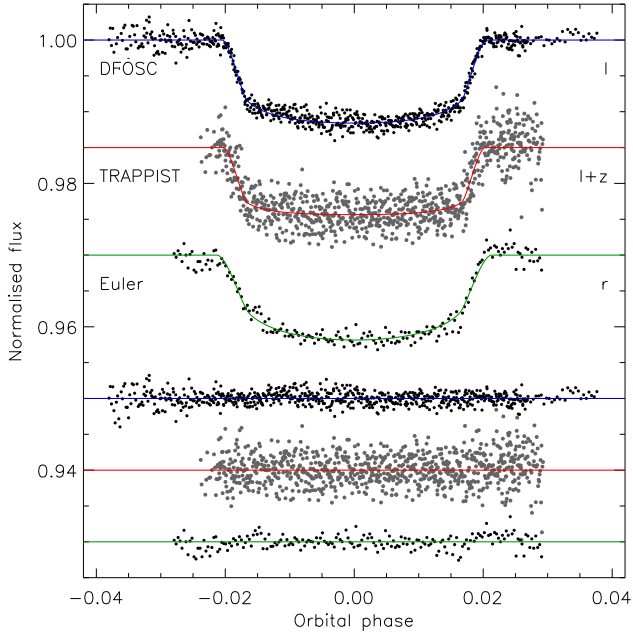}
\caption{\label{fig:22:lcfit} The phased light curves of WASP-22 analysed in
this work, compared to the {\sc jktebop} best fits. The residuals of the fits
are plotted at the base of the figure, offset from unity. Labels give the
source and passband for each dataset. The polynomial baseline functions have
been subtracted from the data before plotting.} \end{figure}

The three DFOSC light curves of WASP-22 were fitted simultaneously (Table\,\ref{tab:lcfit} and Fig.\,\ref{fig:22:lcfit}). A circular orbit was assumed as \citet{Anderson+11aa} found $e < 0.063$ at the 3$\sigma$ level, and \citet{Pont+11mn} found $e < 0.059$ at the 95\% confidence level. The overall quality of the fit is $\chir=1.01$ (remember that the errorbars on each dataset were already scaled to give $\chir=1.0$) which shows that the three light curves give highly consistent transit shapes.

\citet{Anderson+11aa} included in their analysis three new transit light curves of WASP-22, two from the TRAPPIST telescope \citep{Jehin+11msngr} and one from EulerCam on the Euler telescope \citep{Lendl+12aa}. We modelled these in the same way as used for our own data. The follow-up light curve presented by \citet{Maxted+10aj} has only partial coverage of one transit so we did not consider it further in the current work. The full results for WASP-22 are given in Table\,\ref{tab:lcfit} and show an acceptable agreement between the different datasets. For our final values we adopt the weighted means of the individual measurements, calculated by multiplying the probability density functions of the measurements. The final values are consistent with, and an improvement on, previously published values.


\subsection{WASP-41}

\begin{figure} \includegraphics[width=\columnwidth,angle=0]{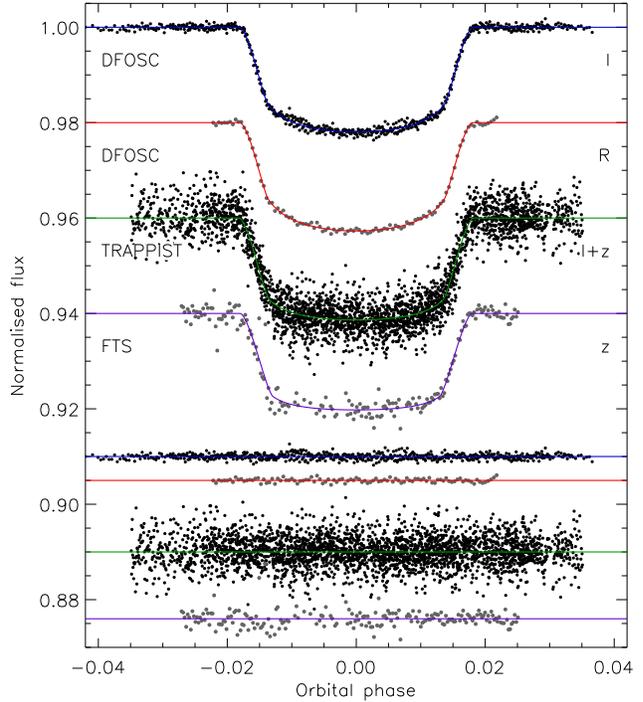}
\caption{\label{fig:41:lcfit} As Fig.\,\ref{fig:22:lcfit}, but for WASP-41.}
\end{figure}

WASP-41 is a trickier \reff{system} because there are anomalies due to starspot-crossing events in at least two of our high-precision light curves. In the current part of the analysis these were ignored, in order to maintain homogeneity of approach, so were therefore basically treated as red noise. A detailed analysis of the spot anomalies will be presented below. Our four DFOSC light curves of WASP-41 were fitted simultaneously (Table\,\ref{tab:lcfit} and Fig.\,\ref{fig:41:lcfit}), for which the best fit returns $\chir = 1.05$ despite the presence of the spot anomalies. A circular orbit was assumed as \citet{Neveu+15xxx} found $e < 0.026$ at the 2$\sigma$ confidence level.

As with WASP-22, the discovery paper of WASP-41 included only one high-precision light curve, which does not cover the full transit. This dataset was not analysed here due to the weak constraints on system properties from light curves missing coverage of the third and fourth contact points in the eclipse. However, \citet{Neveu+15xxx} presented eight new light curves of WASP-41 obtained from three telescopes. The single complete transit from the Danish Telescope, labelled `DFOSC $R$' in Fig.\,\ref{fig:41:lcfit}, and the transit from Faulkes Telescope South (FTS) were each modelled in isolation. The five datasets from TRAPPIST were modelled together, with an extra polynomial to account for the meridian flip at JD 2456402.653 (L.\ Delrez, 2015, priv.\ comm.). The final results are shown in Table\,\ref{tab:lcfit} and agree with, but improve on, published values.


\subsection{WASP-42}

\begin{figure} \includegraphics[width=\columnwidth,angle=0]{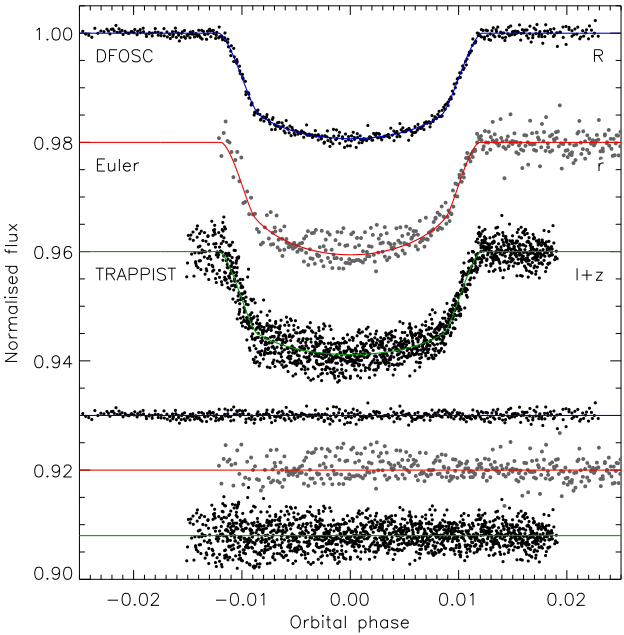}
\caption{\label{fig:42:lcfit} As Fig.\,\ref{fig:22:lcfit}, but for WASP-42.}
\end{figure}

Of the four systems studied in this work, WASP-42 is the only one with an eccentric orbit. \citet{Lendl+12aa} found that their measurement of a small eccentricity was significant at the 99.5\% level. We accounted for this in the {\sc jktebop} modelling by constraining the eccentricity and argument of periastron to be $e = 0.060 \pm 0.013$ and $\omega = 167 \pm 26^\circ$, respectively. \reff{There also appears to be a starspot crossing event just after the midpoint of the third and final transit, which is not surprising given the \Teff\ of the host star. We did not make any attempt to fit this anomaly because it has a low amplitude and no other spot anomalies were seen.}

Our three light curves were all obtained within 34\,d -- the observability of transits in this object has a strong seasonal dependence because its orbital period is close to an integer number of days -- and were modelled together (Table\,\ref{tab:lcfit}). The combined fit has $\chir = 1.09$, once again showing good agreement between our three light curves (Fig.\,\ref{fig:42:lcfit}). We also modelled the light curves from the Euler and TRAPPIST telescopes presented in \citet{Lendl+12aa}, which cover two and four transits respectively. The three light curves are in excellent agreement, with values of \chir\ between 0.30 and 0.57 for the photometric parameters, where \chir\ \reff{is calculated for the individual values of a parameter versus to the weighted mean of the values.}


\subsection{WASP-55}

\begin{figure} \includegraphics[width=\columnwidth,angle=0]{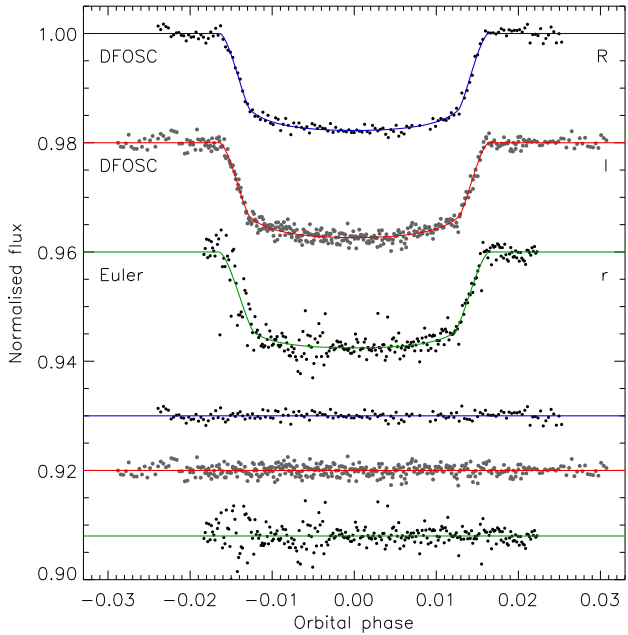}
\caption{\label{fig:55:lcfit} As Fig.\,\ref{fig:22:lcfit}, but for WASP-55.}
\end{figure}

From high-resolution imaging \citet{Evans+15sub} found a faint star close to the WASP-55 system. The star is at an angular distance of $4.345 \pm 0.010$\,arcsec and has a magnitude difference of $5.210 \pm 0.018$ in the $r_{\rm TCI}$ band, which is similar to a combined Gunn $i$+$z$ band. No observations were obtained in the $v_{\rm TCI}$ band, so the colour and therefore spectral type of the faint companion cannot be constrained. We conservatively find that between 50\% and 90\% of the total light from this object is contained in the aperture used for WASP-55 itself, giving a contaminating light fraction of 0.41\% to 0.74\%. We account for this in the {\sc jktebop} fits by setting the third light parameter to be $L_3 = 0.006 \pm 0.003$, where the errorbar has been increased to account for possible differences between the $r_{\rm TCI}$ band used for the high-resolution images and the $R$ and $I$ bands used in the current work.

We observed three transits of WASP-55, one with an $R$ filter and two through an $I$ filter. The two datasets were modelled separately but both with the third light constraint (Fig.\,\ref{fig:55:lcfit}). The results (Table\,\ref{tab:lcfit}) are in good agreement. The discovery paper \citep{Hellier+12mn} presented TRAPPIST light curves of two transits and a Euler light curve of one transit. Both TRAPPIST datasets have only partial coverage of the transit so were not analysed here. The Euler light curve is of decent quality \reff{albeit showing significant red noise,} and was modelled with {\sc jktebop} in the same way as for our own data. As with WASP-22, we obtain weighted means of the photometric parameters by multiplying together the probability density functions of the individual measurements for each parameter. We find \chir\ values less than 1.0 for all photometric parameters, indicating that the results for the different light curves are in good agreement.



\section{Physical properties}

\begin{table} \centering \caption{\label{tab:spec} Spectroscopic properties of the
planet host stars used in the determination of the physical properties of the systems.
\newline {\bf References:} (1) \citet{Mortier+13aa}; (2) \citet{Knutson+14apj};
(3) \citet{Neveu+15xxx}; (4) \citet{Lendl+12aa}; (5) \citet{Hellier+12mn}}
\begin{tabular}{l r@{\,$\pm$\,}l r@{\,$\pm$\,}l r@{\,$\pm$\,}l c}
\hline
Target & \mc{\Teff\ (K)} & \mc{\FeH\ (dex)} & \mc{$K_{\rm A}$ (\ms)} & Refs. \\
\hline
WASP-22 & 6153 & 50 & 0.26 & 0.05 &  70.9 & 1.6 & 1,1,2 \\
WASP-41 & 5546 & 50 & 0.06 & 0.05 & 138   & 2   & 1,1,3 \\
WASP-42 & 5315 & 79 & 0.29 & 0.05 &  64.8 & 1.7 & 1,1,4 \\
WASP-55 & 6070 & 53 & 0.09 & 0.05 &  70   & 4   & 1,1,5 \\
\hline \end{tabular} \end{table}

\begin{table*} \caption{\label{tab:model} Derived physical properties of the four systems.
Where two sets of errorbars are given, the first is the statistical uncertainty and the second is the systematic uncertainty.}
\setlength{\tabcolsep}{4pt}
\begin{tabular}{l l r@{\,$\pm$\,}c@{\,$\pm$\,}l r@{\,$\pm$\,}c@{\,$\pm$\,}l r@{\,$\pm$\,}c@{\,$\pm$\,}l r@{\,$\pm$\,}c@{\,$\pm$\,}l} \hline
Quantity                        & Symbol           & \mcc{WASP-22}               & \mcc{WASP-41}               & \mcc{WASP-42}               & \mcc{WASP-55}               \\
\hline
Stellar mass            (\Msun) & $M_{\rm A}$      & \ermcc{1.249}{0.073}{0.030}{0.015}{0.014}      & 0.987    & 0.021    & 0.026     & 0.951    & 0.037    & 0.051     & \ermcc{1.162}{0.029}{0.033}{0.022}{0.021}      \\
Stellar radius          (\Rsun) & $R_{\rm A}$      & \ermcc{1.255}{0.030}{0.029}{0.005}{0.005}      & 0.886    & 0.009    & 0.008     & 0.892    & 0.021    & 0.016     & \ermcc{1.102}{0.020}{0.015}{0.007}{0.007}      \\
Stellar surface gravity (c.g.s) & $\log g_{\rm A}$ & \ermcc{4.338}{0.027}{0.020}{0.002}{0.002}      & 4.538    & 0.008    & 0.004     & 4.515    & 0.022    & 0.008     & \ermcc{4.419}{0.009}{0.015}{0.003}{0.003}      \\
Stellar density         (\psun) & $\rho_{\rm A}$   & \ercc{0.632}{0.043}{0.041}                     & \mcc{$1.420 \pm 0.034$}         & \mcc{$1.338 \pm 0.092$}         & \ercc{0.869}{0.026}{0.041}                     \\[2pt]
Planet mass             (\Mjup) & $M_{\rm b}$      & \ermcc{0.617}{0.028}{0.017}{0.005}{0.005}      & 0.977    & 0.020    & 0.017     & 0.527    & 0.020    & 0.019     & \ermcc{0.627}{0.037}{0.038}{0.008}{0.007}      \\
Planet radius           (\Rjup) & $R_{\rm b}$      & \ermcc{1.199}{0.046}{0.027}{0.005}{0.005}      & 1.178    & 0.015    & 0.010     & 1.122    & 0.033    & 0.020     & \ermcc{1.335}{0.031}{0.020}{0.008}{0.008}      \\
Planet surface gravity  (\mss ) & $g_{\rm b}$      & \ercc{10.63}{ 0.53}{ 0.71}                     & \mcc{$17.45 \pm  0.46$}         & \mcc{$10.38 \pm  0.61$}         & \ercc{8.73}{0.54}{0.62}                        \\
Planet density          (\pjup) & $\rho_{\rm b}$   & \ermcc{0.334}{0.024}{0.033}{0.001}{0.001}      & 0.558    & 0.020    & 0.005     & 0.349    & 0.029    & 0.006     & \ermcc{0.247}{0.017}{0.021}{0.001}{0.002}      \\[2pt]
Equilibrium temperature (K    ) & \Teq\            & \ercc{1502}{  20}{  20}                        & \mcc{$1242 \pm   12$}           & \mcc{$1021 \pm   19$}           & \ercc{1300}{  15}{  13}                        \\
Orbital semimajor axis  (au   ) & $a$              & \ermcc{0.0489}{0.0010}{0.0004}{0.0002}{0.0002} & 0.0410   & 0.0003   & 0.0004    & 0.0561   & 0.0007   & 0.0010    & \ermcc{0.0558}{0.0005}{0.0005}{0.0004}{0.0003} \\
Age                     (Gyr  ) & $\tau$           & \ermcc{1.3}{0.6}{1.7}{0.4}{0.2}                & \ermcc{1.2}{1.1}{0.0}{0.3}{0.2} & \ermcc{4.4}{3.0}{2.4}{3.2}{2.4} & \ermcc{1.1}{0.8}{0.6}{0.4}{0.1}                \\
\hline \end{tabular} \end{table*}

The results of the above photometric analysis were combined with measured spectroscopic quantities in order to determine the physical properties of the four planetary systems. For each object we used the weighted mean of the measured values of $r_{\rm A}$, $r_{\rm b}$ and $i$ from Table\,\ref{tab:lcfit}. \reff{To these we added} spectroscopic values for the host star's effective temperature, \Teff, metallicity, \FeH, and velocity amplitude, $K_{\rm A}$, from the literature (see Table\,\ref{tab:spec}). These quantities alone are insufficient to yield \reff{the physical properties of the stars or planets}, so the properties of the host stars were \reff{additionally} constrained using tabulated predictions from theoretical models \citep{Claret04aa,Demarque+04apjs,Pietrinferni+04apj,Vandenberg++06apjs,Dotter+08apjs}.

For each object we estimated the value of the velocity amplitude of the planet, $K_{\rm b}$ and calculated the physical properties of the system using this and the measured quantities. We then iteratively adjusted $K_{\rm b}$ to obtain the best agreement between the calculated $\frac{R_{\rm A}}{a}$ and the measured $r_{\rm A}$, and between the \Teff\ and that predicted by the stellar models for the observed \FeH\ and calculated stellar mass ($M_{\rm A}$). This was done for a range of ages, allowing us to identify the overall best fit and age of the system \citep[see][]{Me09mn}. This process was performed for each of the five sets of theoretical models, allowing us to quantify the effect of using theoretical predictions on our results.

The measured physical properties of the four systems are given in Table\,\ref{tab:model}. Statistical errors were calculated by propagating the uncertainties in all the input quantities to each of the output quantities. Systematic uncertainties were obtained by taking the maximum deviation between the final value and the five values from using the different stellar models. Our results are in good agreement with literature values for three of the four systems, but differ in that they are based on more extensive observational data and explicitly account for systematic errors due to the use of theoretical stellar models.

In the case of WASP-22, our measured system properties differ moderately from previous values \citep{Maxted+10aj,Anderson+11aa}. Whilst there are some differences in the photometric parameters from our data, which are of significantly higher quality than the existing TRAPPIST and Euler light curves, the main effect is due to our adoption of the higher and more precise \Teff\ value obtained by \citet{Mortier+13aa} for the host star. For comparison, we calculated an alternative set of results using the lower value of $\Teff = 6020 \pm 50$\,K found from the infrared flux method by \citet{Maxted++11mn}. The mass and radius of the host star change to 1.194\Msun\ and 1.236\Rsun, respectively, and those of the planet to $0.598$\Mjup\ and $1.181$\Rjup. These numbers are all smaller than our adopted values, but in all cases the change is within the errorbars.


\section{Spot modelling of WASP-41}

The third and fourth transits of WASP-41 show clear evidence of starspot activity, manifested as short increases in brightness during transit when the planet crosses areas which are of lower surface brightness than the rest of the stellar photosphere (see Fig.\,\ref{fig:lcall}). The spot crossing events hold information on the size and brightness of the spots, and potentially allow the motion of spots and therefore the rotation of the star to be tracked \citep{Silva08apj,Nutzman+11apj2}. It was for this reason that we observed three transits of WASP-41 over a six-day period in 2015.

\begin{figure}
\includegraphics[width=0.49\columnwidth,angle=0]{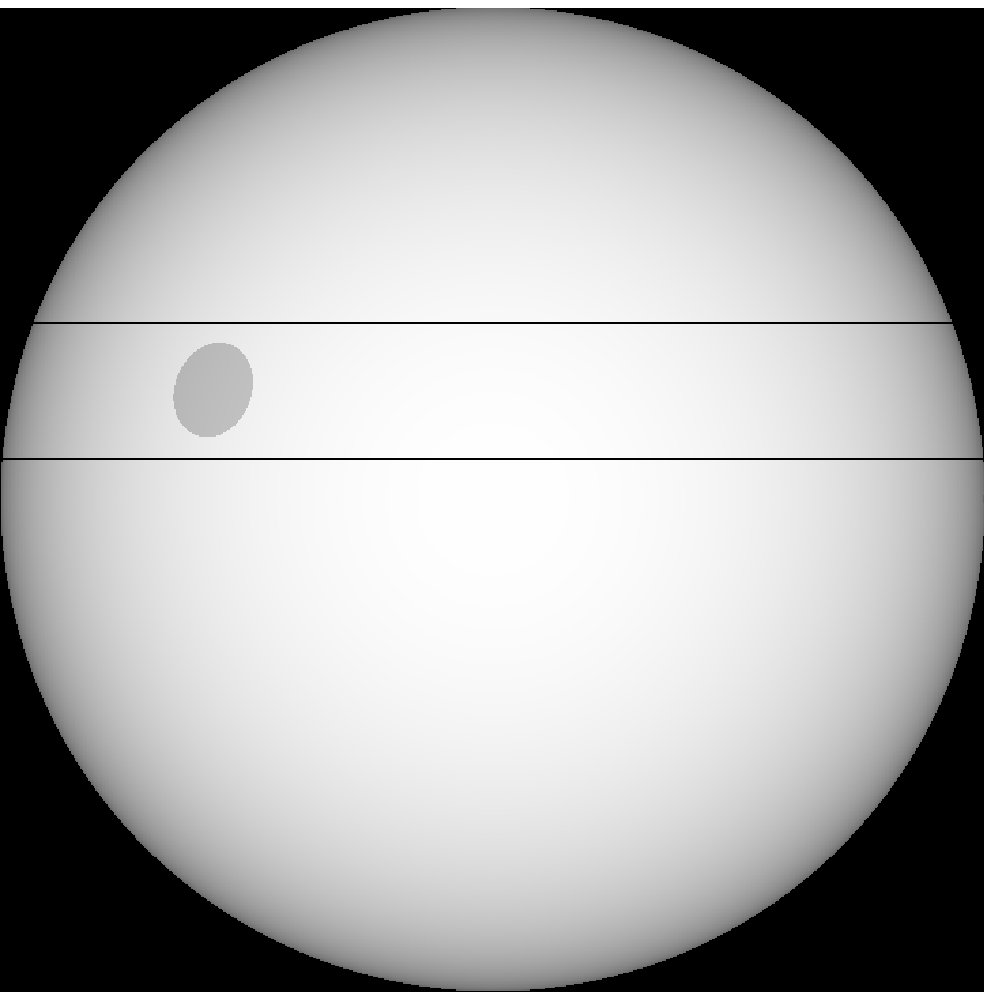}
\includegraphics[width=0.49\columnwidth,angle=0]{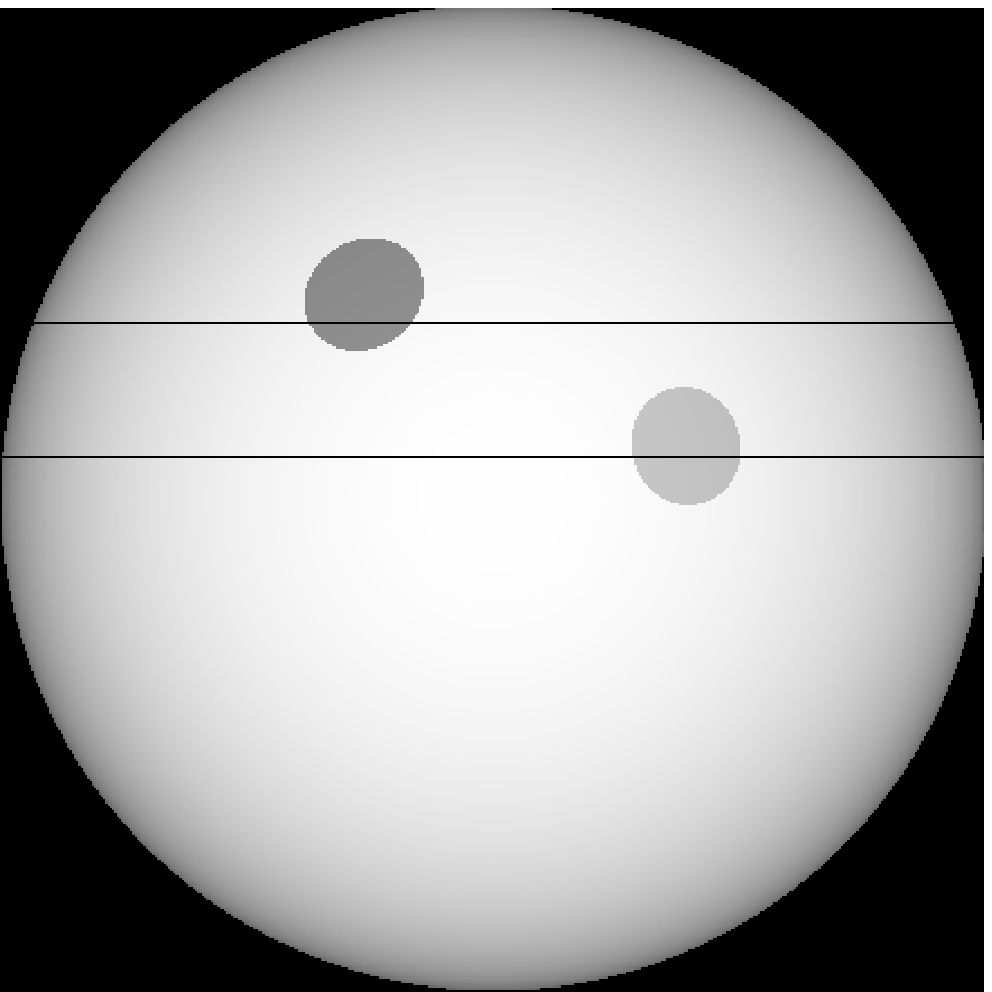}
\caption{\label{fig:41:spotimage} Visualisations of the surface of WASP-41\,A
at the times of the third (left) and fourth (right) transits we observed with
DFOSC, produced by {\sc prism}. The boundaries of the path of the planet are
shown with black lines and the spots are shown with their measured locations,
sizes and contrasts.} \end{figure}

\begin{figure}
\includegraphics[width=\columnwidth,angle=0]{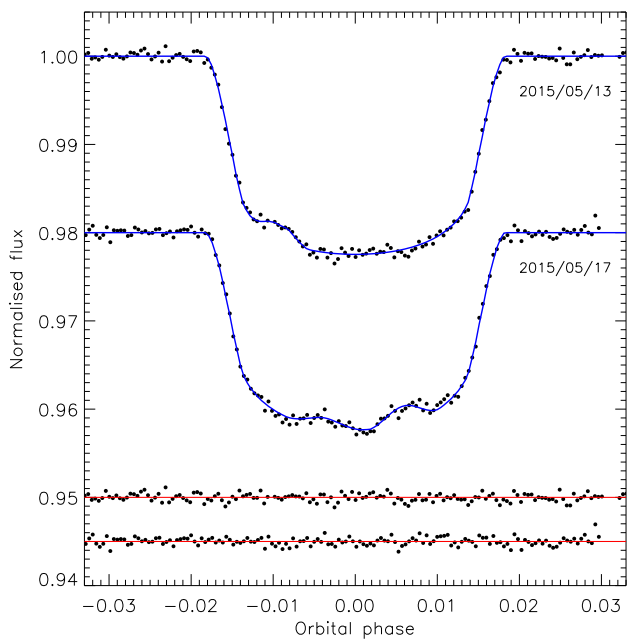}
\caption{\label{fig:41:spotLC} Best fits from {\sc prism}+{\sc gemc} for the
third (upper) and fourth (lower) transits we observed with DFOSC.} \end{figure}

We modelled these two transit light curves using the {\sc prism} and {\sc gemc} codes \citep{Tregloan++13mn,Tregloan++15mn}. {\sc prism} uses a pixellation approach to calculate the light curve of a planet transiting a spotted star, and {\sc gemc} is a hybrid between a Markov chain Monte Carlo and a genetic algorithm to find the best fit to a light curve of a single transit. {\sc gemc} is based on the differential evolution Markov chain approach by \citet{Terbraak06}. The light curve from the night of 2015/05/13 shows clear evidence for one spot crossing -- attempts to fit for a postulated second spot crossing did not lead to a determinate solution -- and the light curve from the night of 2015/05/17 contains two spot crossing events. These two datasets were fitted individually in order to determine the locations, sizes and contrasts of the spots, where `contrast' refers to the ratio of the brightness of the spot to that of the pristine stellar photosphere in the passband \reff{used} to obtain the observations (Figs.\ \ref{fig:41:spotimage} and \ref{fig:41:spotLC}).

\begin{table} \caption{\label{tab:spot} Properties of the spots occulted during two
transits of WASP-41\,A by WASP-41\,b, obtained from modelling the light curves with
{\sc prism}+{\sc gemc}. Longitude and latitude are defined to be zero at the centre
of the stellar disc.}
\begin{tabular}{l r@{\,$\pm$\,}l r@{\,$\pm$\,}l r@{\,$\pm$\,}l}
\hline
Light curve                   & \mc{2015/05/13} & \mc{2015/05/17} & \mc{2015/05/17} \\
Spot number                   & \mc{ }          & \mc{Spot 1}     & \mc{Spot 2}     \\
\hline
Spot longitude ($^\circ$)     & $-$36.3 & 4.5   & $-$37.2 & 2.8   &    23.7 & 1.6   \\
Spot latitude ($^\circ$)      &    15.3 & 10.3  &    27.4 & 6.6   &     8.3 & 6.5   \\
Spot size ($^\circ$)          &    10.4 & 6.5   &    15.5 & 3.5   &    14.3 & 3.2   \\
Spot contrast                 &    0.80 & 0.14  &    0.82 & 0.07  &    0.89 & 0.06  \\
\hline \end{tabular} \end{table}

In order to limit the strong correlation between the orbital inclination (or, equivalently, the impact parameter) of the planet and the latitude of the spot, we fixed $i = 88.7^\circ$ in the {\sc prism}+{\sc gemc} fits. We also fitted for the linear LD coefficient whilst fixing the quadratic coefficient to 0.3. Exploratory fits used a resolution of 15\,pixels for the radius of the planet, for speed, and for final fits we used 50\,pixels to obtain higher precision in the results (see Table\,\ref{tab:spot}).

We found that the latitudes of the spots are not very well determined, as expected for the case where the transit cord of the planet passes close to the centre of the star. Even in the case of a fixed spot contrast, a similar amplitude in flux for the spot crossing event can be obtained for a small spot which is totally occulted by the planet, or a larger spot positioned {\em either above or below the transit cord} which is partially eclipsed by the planet. Similarly, a partially-eclipsed spot can have very little effect on the light curve shape if a modest increase in its size is compensated for by moving its latitude further away from the transit cord, and vice versa. As there is also a known degeneracy between spot contrast and size \citep[e.g.][]{Tregloan++13mn}, but only for those parts of the spot which are occulted by the planet, it is clear that the parameter space for spot-modelling is inherently complex and degenerate.


However, the spot longitudes are very well determined by our data (see Table\,\ref{tab:spot}) because they govern the times at which spot crossings are detected. We now assert that the spot observed on 2015/05/13 is the same as the second spot observed on 2015/05/17. In this case the change in the longitude and latitude of the spot (both defined to be zero at the centre of the stellar disc) are $60.0 \pm 4.8^\circ$ and $-7 \pm 12^\circ$, respectively, and the time difference between the midpoints of the two spot crossings is $3.097 \pm 0.005$\,d. If the spot moved directly from the first to the second location, this gives a rotation period of $18.6 \pm 1.5$\,d at a latitude of approximately $12^\circ$\footnote{Alternative assumptions all yield much shorter rotation periods. For example, if the spot moves from the first to the second location but in a retrograde direction it has to travel 300$^\circ$ and the rotation period is $3.7 \pm 0.3$\,d. If the rotation is instead prograde but the star rotates once plus the 60$^\circ$ difference in longitude, then the rotation period is $2.4 \pm 0.2$\,d. All other possibilities require additional rotations of the star between the two detections of the spot, and thus yield ever-shorter rotation periods.}, in good agreement with the value of $18.41 \pm 0.05$\,d measured from the spot-induced brightness modulation of WASP-41\,A by \citet{Maxted+11pasp}. This equates to a projected rotational velocity of $v \sin i = 2.4 \pm 0.2$\kms, again in good agreement with the measurements from \citet{Maxted+11pasp} and \citet{Neveu+15xxx}. We also note that the measured radii and contrasts of the two spot events agree to within the (relatively large) errorbars.

The assumption that we have detected the same spot twice in two different positions on the stellar surface leads directly to a detection of a change of latitude and therefore a measurement of the sky-projected orbital obliquity of the system of $\lambda = 6 \pm 11^\circ$. \citet{Neveu+15xxx} used spectroscopy during a transit of WASP-41 to measure $\lambda = \er{29}{10}{14}$$^\circ$ with, and $\lambda = 48 \pm 29^\circ$ without, a Bayesian prior on the projected rotational velocity. Our revised value is consistent with both to within 1.3$\sigma$, removes the ambiguity due to choice of methodology, and also banishes the previous hints of orbital misalignment. WASP-41 is therefore another example of an aligned system containing a transiting hot Jupiter and a cool star with a precisely measured projected orbital obliquity.


\section{Transit timing analysis}                                                                                                    \label{sec:porb}

\begin{table*} \begin{center}
\caption{\label{tab:minima} Times of minimum light and their residuals versus the ephemerides derived in this work.}
\begin{tabular}{l l l r r l} \hline
Target & Time of minimum & Uncertainty & Cycle & Residual & Reference \\
       & (BJD/TDB)) & (d) & number & (d) &       \\
\hline
WASP-22 & 2454041.91397 & 0.00122 & $-$422.0 & $-$0.00146 & This work (WASP-South 2006) \\   
WASP-22 & 2454409.31798 & 0.00078 & $-$318.0 & $-$0.00143 & This work (WASP-South 2007) \\   
WASP-22 & 2455518.59851 & 0.00059 &   $-$4.0 & $ $0.00167 & This work (TRAPPIST)  \\   
WASP-22 & 2455532.72730 & 0.00038 &      0.0 & $-$0.00046 & This work (Euler)     \\   
WASP-22 & 2455532.72761 & 0.00053 &      0.0 & $-$0.00015 & This work (TRAPPIST)  \\   
WASP-22 & 2455896.60029 & 0.00081 &    103.0 & $ $0.00128 & This work (84\,cm)      \\   
WASP-22 & 2456189.81595 & 0.00022 &    186.0 & $ $0.00029 & This work (Danish)     \\   
WASP-22 & 2456560.75224 & 0.00029 &    291.0 & $-$0.00014 & This work (Danish)     \\   
WASP-22 & 2457270.83104 & 0.00022 &    492.0 & $-$0.00020 & This work (Danish)     \\   
\hline
WASP-41 & 2454201.86515 & 0.00180 & $-$588.0 & $-$0.00201 & This work (WASP-South 2007)                     \\   
WASP-41 & 2454549.83845 & 0.00100 & $-$474.0 & $-$0.00248 & This work (WASP-South 2008)                     \\   
WASP-41 & 2455642.60026 & 0.00027 & $-$116.0 & $-$0.00043 & This work (TRAPPIST)                      \\   
WASP-41 & 2455654.81018 & 0.00020 & $-$112.0 & $-$0.00012 & This work (TRAPPIST)                      \\   
WASP-41 & 2455663.96815 & 0.00024 & $-$109.0 & $ $0.00065 & This work (FTS)                           \\   
WASP-41 & 2455694.49015 & 0.00091 &  $-$99.0 & $-$0.00137 & This work (TRAPPIST)                      \\   
WASP-41 & 2455725.01389 & 0.00045 &  $-$89.0 & $-$0.00164 & Tan (ETD)                                 \\   
WASP-41 & 2455947.84052 & 0.00021 &  $-$16.0 & $-$0.00032 & This work (84\,cm)                          \\   
WASP-41 & 2455996.67838 & 0.00032 &      0.0 & $-$0.00089 & This work (TRAPPIST)                      \\   
WASP-41 & 2456402.65000 & 0.00021 &    133.0 & $ $0.00132 & This work (TRAPPIST)                      \\   
WASP-41 & 2456402.64891 & 0.00007 &    133.0 & $ $0.00023 & This work (Danish)                         \\   
WASP-41 & 2456424.01544 & 0.00030 &    140.0 & $-$0.00005 & Tan (ETD)                                 \\   
WASP-41 & 2456698.73202 & 0.00038 &    230.0 & $ $0.00040 & Masek (ETD)                               \\   
WASP-41 & 2456765.88379 & 0.00093 &    252.0 & $-$0.00067 & Evans (ETD)                               \\   
WASP-41 & 2456768.93761 & 0.00041 &    253.0 & $ $0.00075 & Evans (ETD)                               \\   
WASP-41 & 2456808.61760 & 0.00009 &    266.0 & $-$0.00048 & This work (Danish)                         \\   
WASP-41 & 2456820.82714 & 0.00058 &    270.0 & $-$0.00055 & Evans (ETD)                               \\   
WASP-41 & 2457153.53964 & 0.00009 &    379.0 & $ $0.00019 & This work (Danish)                         \\   
WASP-41 & 2457156.59146 & 0.00015 &    380.0 & $-$0.00039 & This work (Danish, {\sc prism}+{\sc gemc}) \\   
WASP-41 & 2457159.64436 & 0.00017 &    381.0 & $ $0.00011 & This work (Danish, {\sc prism}+{\sc gemc}) \\   
\hline
WASP-42 & 2454554.59925 & 0.00140 & $-$220.0 & $ $0.00199 & This work (WASP-South 2008) \\   
WASP-42 & 2455625.65818 & 0.00027 &   $-$5.0 & $-$0.00070 & This work (TRAPPIST) \\   
WASP-42 & 2455630.64068 & 0.00029 &   $-$4.0 & $ $0.00012 & This work (TRAPPIST) \\   
WASP-42 & 2455650.56811 & 0.00033 &      0.0 & $ $0.00082 & This work (TRAPPIST) \\   
WASP-42 & 2455655.54875 & 0.00032 &      1.0 & $-$0.00022 & This work (TRAPPIST) \\   
WASP-42 & 2455645.58567 & 0.00035 &   $-$1.0 & $ $0.00007 & This work (Euler) \\   
WASP-42 & 2456437.67306 & 0.00014 &    158.0 & $ $0.00004 & This work (Danish) \\   
WASP-42 & 2456462.58139 & 0.00010 &    163.0 & $-$0.00004 & This work (Danish) \\   
WASP-42 & 2456472.54492 & 0.00018 &    165.0 & $ $0.00012 & This work (Danish) \\   
\hline
WASP-55 & 2453902.56555 & 0.00340 & $-$563.0 & $-$0.00093 & This work (WASP-South 2006) \\   
WASP-55 & 2454201.76435 & 0.00280 & $-$496.0 & $ $0.00073 & This work (WASP-South 2007) \\   
WASP-55 & 2454581.34095 & 0.00220 & $-$411.0 & $-$0.00115 & This work (WASP-South 2008) \\   
WASP-55 & 2454951.98517 & 0.00210 & $-$328.0 & $-$0.00415 & This work (WASP-South 2009) \\   
WASP-55 & 2455309.24157 & 0.00170 & $-$248.0 & $ $0.00192 & This work (WASP-South 2010) \\   
WASP-55 & 2455715.61277 & 0.00046 & $-$157.0 & $ $0.00088 & This work (Euler) \\   
WASP-55 & 2456416.71548 & 0.00016 &      0.0 & $-$0.00017 & This work (Danish) \\   
WASP-55 & 2456778.43544 & 0.00283 &     81.0 & $ $0.00383 & Lomoz (ETD) \\   
WASP-55 & 2456778.42973 & 0.00200 &     81.0 & $-$0.00188 & Lomoz (ETD) \\   
WASP-55 & 2456827.55355 & 0.00023 &     92.0 & $ $0.00002 & This work (Danish) \\   
WASP-55 & 2457135.68202 & 0.00019 &    161.0 & $ $0.00009 & This work (Danish) \\   
\hline \end{tabular} \end{center} \end{table*}

\begin{figure*}
\includegraphics[width=\textwidth,angle=0]{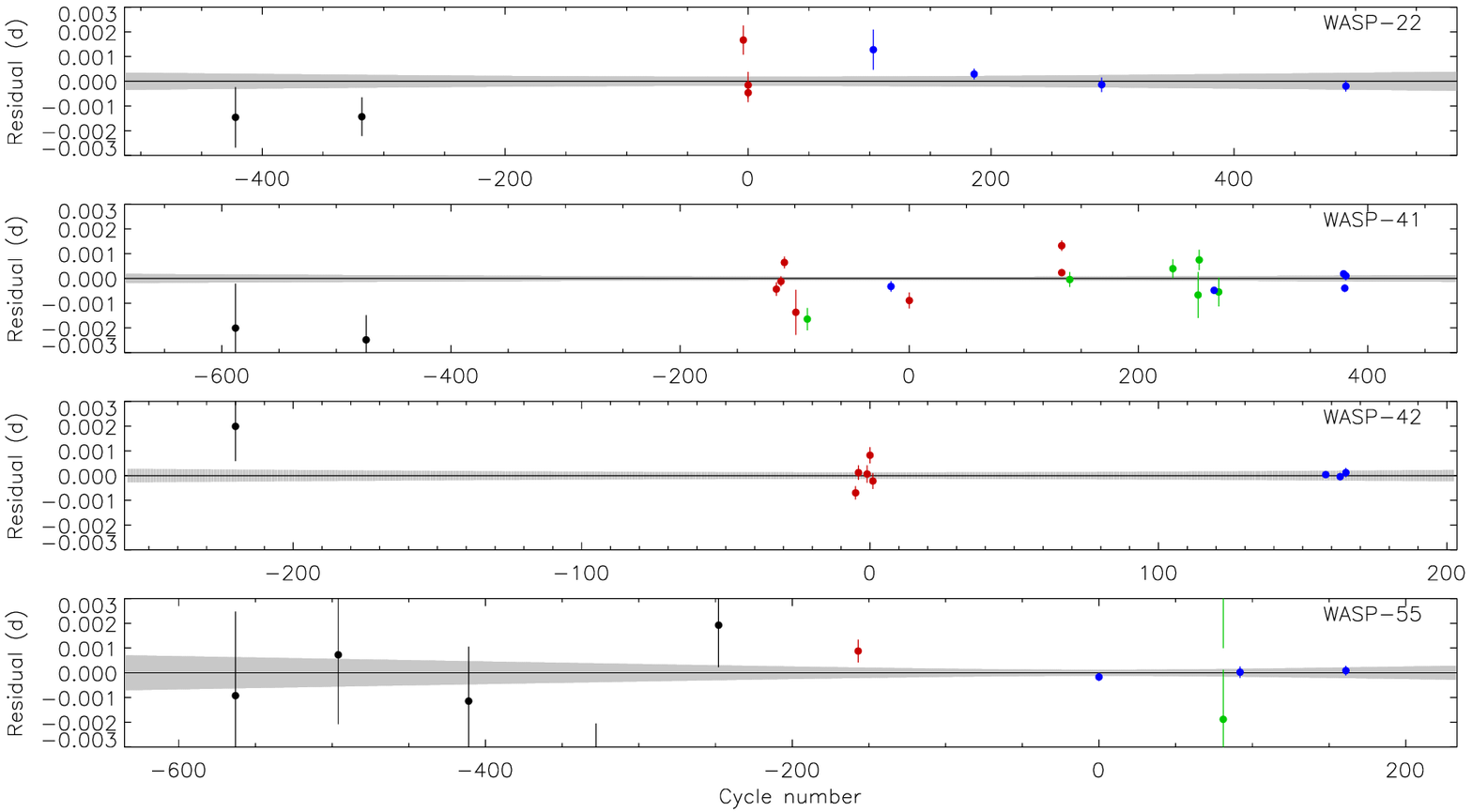}
\caption{\label{fig:minima} Plot of the residuals of the timings of mid-transit
versus a linear ephemeris. The results from this work are shown in blue and from
amateur observers in green. Our reanalysis of published data are shown in black
for WASP-South observations and in red for other sources. The grey-shaded regions
show the 1$\sigma$ uncertainty in the ephemeris as a function of cycle number.}
\end{figure*}

Each of the transit light curves available for the four systems were fitted with the {\sc jktebop} code in order to determine the time of midpoint of the transit. We did not apply this analysis to light curves lacking complete coverage of a transit, as these give noisy and possibly biased values \citep[e.g.][]{Gibson+09apj}. We also obtained the WASP-South light curves, divided them into individual observing seasons, and fit each season separately to obtain a time of minimum close to the midpoint of the data. In the case of the WASP-South data, which have a very high scatter compared to the follow-up light curves, we fixed the values of the photometric parameters to the best estimates obtained in Section\,\ref{sec:lc} and thus fitted for only the time of midpoint and the out-of-transit brightness of the system. All measured transit times were moved to the TDB timescale.

We also included times of minimum for WASP-41 and WASP-55 from the Exoplanet Transit Database\footnote{The Exoplanet Transit Database (ETD) can be found at: {\tt http://var2.astro.cz/ETD/credit.php}} \citep{Poddany++10newa}, which provides data and transit times from amateur observers affiliated with TRESCA\footnote{The TRansiting ExoplanetS and CAndidates (TRESCA) website can be found at: {\tt http://var2.astro.cz/EN/tresca/index.php}}. We assumed that the times were on the UTC timescale and converted them to TDB.

For each object we fitted the times of mid-transit with straight lines to determine a new linear orbital ephemeris. Table\,\ref{tab:minima} gives all transit times plus their residuals versus the fitted ephemeris. In cases where the \chir\ was greater than 1.0 we scaled the uncertainties to give $\chir = 1.0$. $E$ gives the cycle count versus the reference epoch, and the bracketed numbers show the uncertainty in the final digit of the preceding number.

The new ephemeris for WASP-22 is based on eight timing measurements and is:
$$ T_0 = {\rm BJD(TDB)} \,\, 2\,455\,532.72776 (22) \, + \, 3.53273064 (70) \times E $$
where the fit has $\chir = 1.46$. The timebase of the ephemeris was chosen to be close to the weighted mean of the data and coincides with the transit observed simultaneously by the TRAPPIST and Euler telescopes. The most discrepant timing is the measurement from TRAPPIST data at cycle $-4$, which differs by 2.4$\sigma$ from the timing predicted by the ephemeris. This is not sufficient grounds to reject the datapoint, so we did not do so. Instead, the errorbar for the ephemeris were multiplied by $\sqrt{1.46}$ to account for the excess \chir.

For WASP-41 we have 19 timings and obtain the ephemeris:
$$ T_0 = {\rm BJD(TDB)} \,\, 2\,455\,996.67927 (10) \, + \, 3.05240154 (41) \times E $$
with $\chir = 2.67$. This is the largest \chir\ among the four objects in this work, and occurs for the system with the most active host star. The errorbars of the ephemeris have been inflated to account for the excess \chir. The times of midpoint of the final two DFOSC transits were obtained using {\sc prism}+{\sc gemc}, which agree with the midpoints obtained using {\sc jktebop} to within 0.00015\,d. This is in line with expectations for the effects of starspots \citep{Barros+13mn,Oshagh+13aa,Ioannidis+16aa}.

For WASP-42 there are three timings from data in the current work, five from published follow-up light curves and one from WASP-South observations in the 2008 season. Whilst there are plenty of WASP-South observations from 2006 and 2007, there is no coverage of transits due to the near-integer orbital period of the system. We find the ephemeris:
$$ T_0 = {\rm BJD(TDB)} \,\, 2\,455\,650.56728 (15) \, + \, 4.9816819 (11) \times E $$
where $\chir = 1.35$ and the errorbars have been inflated to account for this.

For WASP-55 we have 11 timings which yield this ephemeris:
$$ T_0 = {\rm BJD(TDB)} \,\, 2\,456\,416.71565 (13) \, + \, 4.4656291 (11) \times E $$
with $\chir = 1.10$ (accounted for in the errorbars).

Fig.\,\ref{fig:minima} shows the residuals versus the linear ephemeris for each of our four targets. No transit timing variations are apparent, and there are too few timing measurements for a search for such variations to be useful. Our period values for all four systems are consistent with previous measurements but are significantly more precise due to the addition of new high-quality data and a longer temporal baseline.


\section{Summary and conclusions}                                                                                                 \label{sec:summary}

\begin{figure} \includegraphics[width=\columnwidth,angle=0]{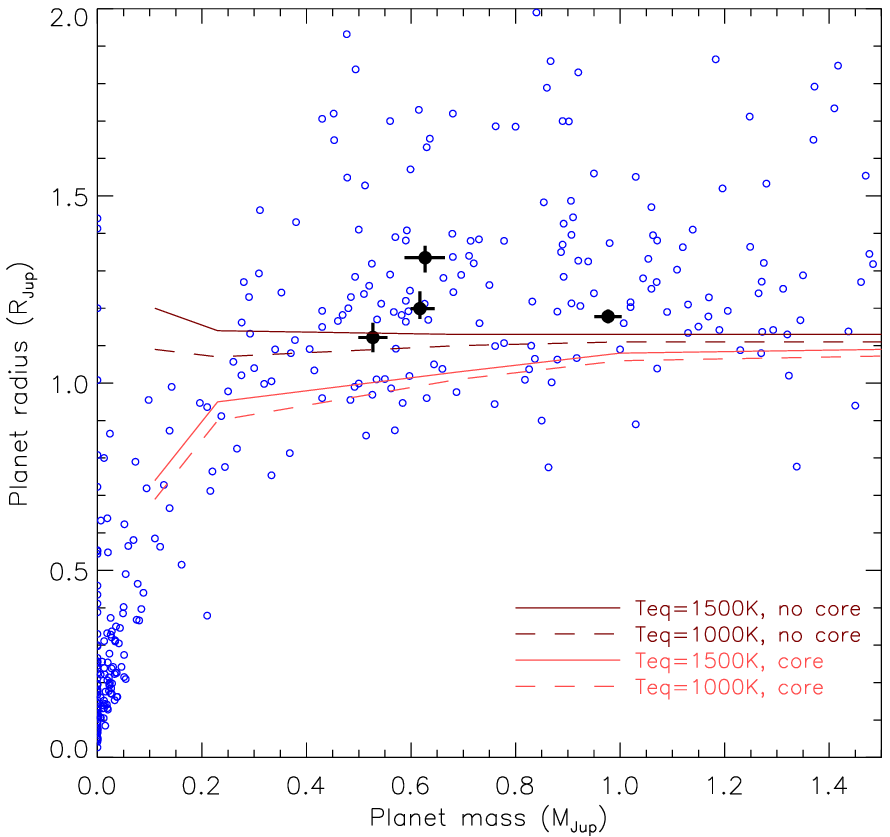}
\caption{\label{fig:m2r2bodenheimer} Plot of planet radii versus their masses.
WASP-22\,b, WASP-41\,b, WASP-42\,b and WASP-55\,b are indicated using black filled
circles. The overall population of planets is shown using blue open circles without
errorbars, using data taken from TEPCat on 2015/11/09. The lines show the predicted
planet radii of gas-giants from \citet[][their table\,1]{Bodenheimer++03apj} for
two different equilibrium temperatures (1000\,K and 1500\,K) which bracket the four
planets, and for with and without a solid 20\Mearth\ core (see key).} \end{figure}

\begin{figure} \includegraphics[width=\columnwidth,angle=0]{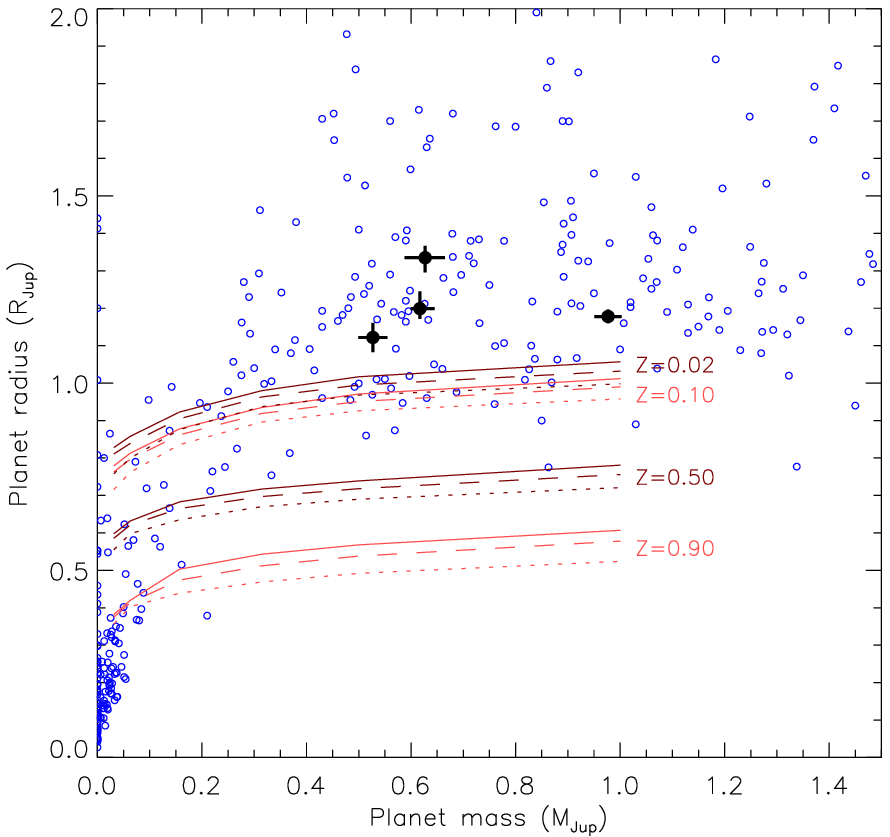}
\caption{\label{fig:m2r2baraffe} As Fig.\,\ref{fig:m2r2bodenheimer} but with
different theoretical predictions. The lines show the predicted planet radii
of gas-giants from \citet[][their table\,4]{Baraffe++08aa} for four different
heavy-element mass fractions $Z$, labelled on the plot. The unbroken lines
show predictions for a planet of age 0.5\,Gyr, the dashed lines for an age of
1\,Gyr and the dotted lines for an age of 5\,Gyr. The lines are colour-coded
for clarity.} \end{figure}

WASP-22, WASP-41, WASP-42 and WASP-55 are four systems containing transiting hot Jupiters with large radii. We have presented high-precision photometry of 13 transits of the four systems, and used these data to refine their measured orbital ephemerides and physical properties. Our light curves of each system contain highly consistent transit shapes, demonstrating the repeatability of observations taken with the telescope-defocussing method. We find no evidence for transit timing variations in any of the systems, and our new measurements of the system properties are mostly in good agreement with previous determinations based on fewer and less precise data.

All four planets have inexplicably larger radii than expected from theoretical models \citep{Bodenheimer++03apj,Fortney++07apj,Baraffe++08aa}. Fig.\,\ref{fig:m2r2bodenheimer} compares the positions of the four planets in the mass--radius diagram to the overall sample of planets\footnote{Data on planetary systems were taken from TEPCat on 2015/11/09. Figs.\ \ref{fig:m2r2bodenheimer} and \ref{fig:m2r2baraffe} show only those planets regarded as ``well-studied'', i.e.\ excluding \reff{planets which have been characterised as part of a large sample of planetary systems without receiving significant individual attention}.} and to predictions from \citet{Bodenheimer++03apj} for planetary equilibrium temperatures similar to those for the four planets which are the subject of the current work. Whilst WASP-22\,b is well represented by models without a heavy-element core, the other three planets are significantly larger than predicted even for coreless gas giants. For comparison, Fig.\,\ref{fig:m2r2baraffe} shows the same mass--radius diagram but with the predictions of the \citet{Baraffe++08aa} theoretical models for a range of heavy-element mass fractions, $Z$. All four planets are larger than model predictions even for $Z = 0.02$, which yields the largest planetary radii of all the model sets. It is clear that all four planets are more inflated than expected, particularly WASP-55, and are therefore good candidates for the characterisation of their atmospheres via transmission spectroscopy and photometry \citep[e.g.][]{Nikolov+14mn,Mallonn+15aa}.

\begin{figure*} \includegraphics[width=\textwidth,angle=0]{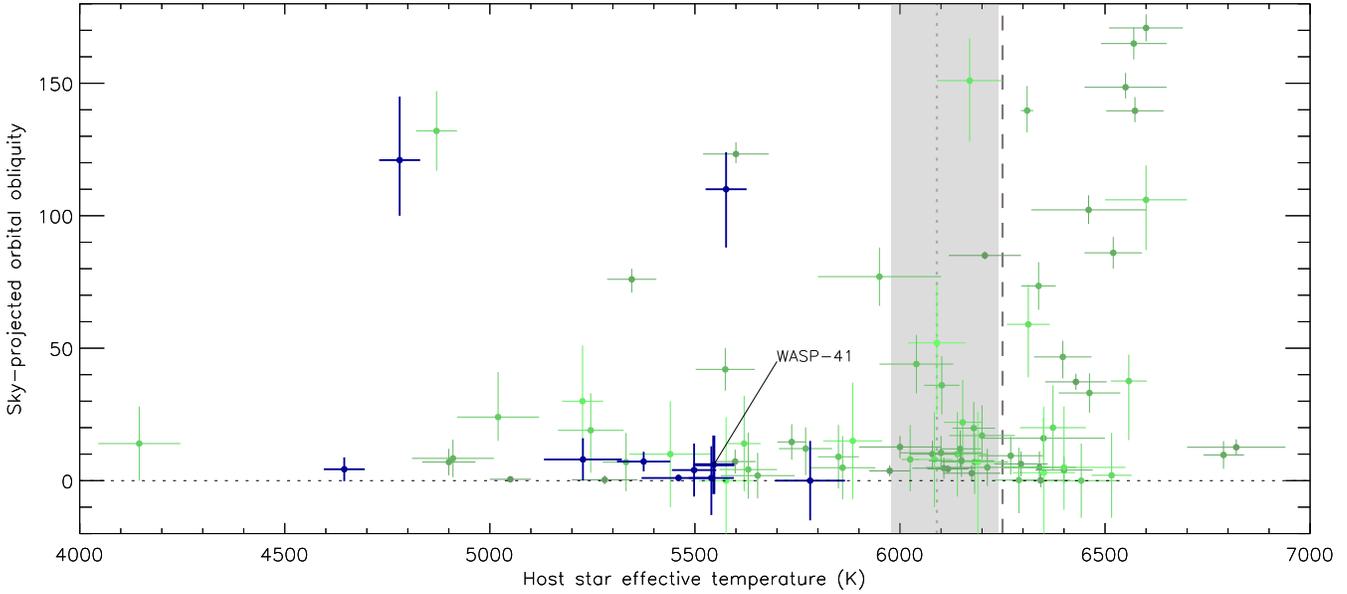}
\caption{\label{fig:tefflambda} Plot of the sky-projected orbital obliquity angle $\lambda$ versus
the \Teff\ of the host star. The data come from TEPCat, obtained on 2015/11/09, and only the best
value for each planetary system is plotted. Green lines show the measurements, graded for clarity
from dark to light green depending on the size of the errorbars. Measurements obtained from spot
tracking are shown with thicker blue lines. The \Teff\ values proposed as boundaries are shown
using a grey dashed line \citep[6250\,K;][]{Winn+10apj3} and a grey shaded region with a dotted
line to indicate the value and its uncertainties \citep[\er{6090}{150}{110}\,K;][]{Dawson14apj}.
Published values of $\lambda$ have been adjusted by $\pm$180$^\circ$ to bring them into the
interval [$0^\circ$,$180^\circ$] \citep[see][]{CridaBatygin14aa}.} \end{figure*}

\begin{table*} \begin{center}
\caption{\label{tab:spotlambda} Published measurements of the sky-projected,
$\lambda$, and true, $\psi$, orbital obliquities obtained from spot-tracking analyses.}
\begin{tabular}{l c c c l}
\hline
System & Host star \Teff\ (K) & $\lambda$ ($^\circ$) & $\psi$ ($^\circ$) & Reference \\
\hline
HAT-P-11  & $4780 \pm 50$ &\er{105}{16}{12} or \er{121}{24}{21}&\er{106}{15}{11} or \er{97}{8}{4}&\citet{SanchisWinn11apj}\\
HATS-02   & $5227 \pm 95$ & $8 \pm 8$                          &                                 & \citet{Mohler+13aa}    \\
\reff{Kepler-17} & $5781 \pm 85$ & $0 \pm 15$                         &                                 & \reff{\citet{Desert+11apjs}}  \\
Kepler-30 & $5498 \pm 54$ & $-1 \pm 10$ or $4 \pm 10$          &                                 & \citet{Sanchis+12nat}  \\
Kepler-63 & $5576 \pm 50$ & \er{-110}{22}{14}                  & \er{145}{9}{14}                 & \citet{Sanchis+13apj2} \\
Qatar-2   & $4645 \pm 50$ & $4.3 \pm 4.5$                      &                                 & \citet{Mancini+14mn}   \\
WASP-4    & $5540 \pm 55$ & \er{-1}{14}{12}                    &                                 & \citet{Sanchis+11apj}  \\
WASP-6    & $5375 \pm 65$ & $7.2 \pm 3.7$                      &                                 & \citet{Tregloan++15mn} \\
WASP-19   & $5460 \pm 90$ & $1.0 \pm 1.2$                      &                                 & \citet{Tregloan++13mn} \\
WASP-41   & $5546 \pm 33$ & $6 \pm 11$                         &                                 & This work              \\
\hline \end{tabular} \end{center} \end{table*}

Two of our transit light curves of WASP-41 show clear evidence for spot activity, with one spot crossing event observed on 2015/05/13 and two on 2015/05/17. We make the assumption that the spot observed on 2015/05/13 is the same as the second spot observed on 2015/05/17, as the measured spot radii and contrasts agree, and the resulting stellar rotation period and velocity are highly consistent with previous measurements obtained using different methods. The change in longitude of the spot then gives a stellar rotation period of $18.6 \pm 1.5$\,d and $v \sin i$ of $2.4 \pm 0.2$\kms, both at a latitude of approximately $12^\circ$. The change in latitude yields a measurement of the sky-projected orbital obliquity of the system of $\lambda = 6 \pm 11^\circ$, which is significantly more precise than a previous measurement obtained via the Rossiter-McLaughlin effect.

Spectroscopic measurements of $\lambda$ are notoriously difficult for cool stars because the amplitude of the Rossiter-McLaughlin effect depends on the $v \sin i$ of the host star, which is typically very low below \Teff\ values of roughly 5500\,K. Starspot tracking is a major contributor in this domain, with a total of ten determinations to date (Table\,\ref{tab:spotlambda} and Fig.\,\ref{fig:tefflambda}). The host stars have \Teff\ values between 4645\,K and 5576\,K, so are all cool stars where Rossiter-McLaughlin measurements are difficult \citep[e.g.][]{Albrecht+11apj2} but stars often show strong spot activity. The orbital obliquity of systems containing cool stars is a useful probe of tidal theory \citep[e.g.][]{Esposito+14aa,Mancini+15aa}, and a statistically significant sample of such measurements is both scientifically important and observationally tractable via starspot tracking analyses such as that performed for WASP-41.


\section*{Acknowledgements}

%
The operation of the Danish 1.54m telescope is financed by a grant to UGJ from the Danish Council for Independent Research, Natural Sciences (FNU).
The reduced light curves presented in this work will be made available at the CDS ({\tt http://vizier.u-strasbg.fr/}) and at {\tt http://www.astro.keele.ac.uk/jkt/}.
We thank Laetitia Delrez and Marion Neveu-VanMalle for providing published light curves of WASP-41.
J\,Southworth acknowledges financial support from the Leverhulme Trust in the form of a Philip Leverhulme Prize.
JTR acknowledges financial support from ORAU (Oak Ridge Associated Universities) and NASA in the form of a NASA Post-Doctoral Programme (NPP) Fellowship.
DFE is funded by the UK's Science and Technology Facilities Council.
EU-S acknowledges the support of CONICYT QUIMAL 130004 project.
%
Funding for the Stellar Astrophysics Centre in Aarhus is provided by The Danish National Research Foundation (grant agreement no.\ DNRF106). The research is supported by the ASTERISK project (ASTERoseismic Investigations with SONG and Kepler) funded by the European Research Council (grant agreement no.\ 267864).
%
%
TCH acknowledges KASI research grants \#2012-1-410-02, \#2013-9-400-00, \#2014-1-400-06 and \#2015-1-850-04.
NP acknowledges funding by the Gemini-Conicyt Fund, allocated to project No.\ 32120036.
GD acknowledges Regione Campania for support from POR-FSE Campania 2014--2020.
%
%
YD, AE, OW and J\,Surdej acknowledge support from the Communaut\'e fran\c{c}aise de Belgique - Actions de recherche concert\'ees - Acad\'emie Wallonie-Europe.
The following internet-based resources were used in research for this paper: the ESO Digitized Sky Survey; the NASA Astrophysics Data System; the SIMBAD database and VizieR catalogue access tool operated at CDS, Strasbourg, France; and the ar$\chi$iv scientific paper preprint service operated by Cornell University.

\bibliographystyle{mn_new}


\end{document}